\documentclass[times, review,nopreprintline]{elsarticle}

\usepackage{medima}
\usepackage{framed,multirow}

\usepackage{amssymb}
\usepackage{latexsym}

\usepackage{amsmath}
\usepackage{makecell}
\usepackage{longtable}

\definecolor{newcolor}{rgb}{.8,.349,.1}

\begin{document}


\begin{frontmatter}

\title{Graph Convolutional Networks in Multi-modality Medical Imaging: Concepts, Architectures, and Clinical Applications \tnoteref{tnote1}}%

\author[1]{Kexin \snm{Ding}}
\author[2]{Mu \snm{Zhou}}
\author[3,4]{Zichen \snm{Wang}}
\author[5]{Qiao \snm{Liu}}
\author[3,4]{Corey W. \snm{Arnold}}
\author[6]{Shaoting \snm{Zhang}\corref{cor1}}
\author[7]{Dimitri N. \snm{Metaxas}\corref{cor1}}
\cortext[cor1]{Corresponding author: 
  Tel.: +1-848-445-2914;
  fax: ++1-848-445-2914;}
\ead{dnm@cs.rutgers.edu}

\address[1]{Department of Computer Science, UNC Charlotte, Charlotte, NC, USA}
\address[2]{Sensebrain Research, San Jose, USA}
\address[3]{Computational Diagnostics Lab, UCLA, Los Angeles, USA}
\address[4]{Department of Bioengineering, UCLA, Los Angeles, USA}
\address[5]{Department of Statistics, Stanford University, Stanford, USA}
\address[6]{Shanghai Artificial Intelligence Laboratory, Shanghai, China}
\address[7]{Department of Computer Science, Rutgers University, New Jersey, USA }

\received{xxx}
\finalform{xxx}
\accepted{xxx}
\availableonline{xxx}
\communicated{xxx}

\begin{abstract}
Image-based characterization and disease understanding involve integrative analysis of morphological, spatial, and topological information across biological scales. The recent surge of graph convolutional networks (GCNs) provides opportunities to address this information complexity via graph-driven architectures. GCNs have demonstrated their capability to perform feature aggregation, interaction, and reasoning with remarkable flexibility and efficiency. These advances of GCNs have prompted a new wave of research and application in medical imaging analysis with an overarching goal of improving quantitative disease understanding, monitoring, and diagnosis. Yet daunting challenges remain for designing the important image-to-graph transformation for multi-modality medical imaging and gaining insights into model interpretation and enhanced clinical decision support. In this review, we share rapid developments of GCNs in the context of medical image analysis including radiology,  histopathology, and other related imaging modalities. We discuss the fast-growing synergy of graph network architectures and medical imaging components to advance our assessment of disease status and outcome in clinical tasks. To provide a guideline to foster cross-disciplinary research, we present emerging opportunities and identify common challenges in image-based GCNs and their extensions in model interpretations, technical advancements, large-scale benchmarks to transform the scope of medical image studies and related graph-driven medical research.

\end{abstract}

%
%

\end{frontmatter}

\section{Introduction}
Graph representation has been broadly studied in information extraction, relational representation, and multi-modality data fusion~\citep{RN1,RN2,RN3}. The rich topological and spatial characteristics of graphs essentially uncover differential relations among individual graph elements~\citep{RN4}. In medical image analysis, the diverse shape, anatomy, and appearance information provide a key data source to characterize the interactions among the diagnostic region of interests (ROIs) and reveal disease status~\citep{RN5}. Therefore, image-based graph modeling and inference can deepen our understanding of the complex relational patterns hidden in disease tissue regions. The surge of graph convolutional networks (GCNs), a branch of deep learning characterized by graph-level model development, has brought a new wave of information fusion techniques through their widespread applications in medical imaging, from disease classification~\citep{RN6}, tumor segmentation~\citep{RN7}, to patient outcome prediction~\citep{RN8}. 

Graph convolutional networks (GCNs) explore the heterogeneous graph data via a series of graph-level convolution, sampling, and enabling the model inference on both graph node attributes and relational structures~\citep{RN4}. The development of GCNs extends conventional graph embedding methods (e.g., Deepwalk~\citep{RN9}  and node2vec~\citep{RN10}) primarily on generating a low-dimensional graph representation without considering node attributes. To enable a synergistic analysis with medical imaging, GCNs demonstrate their multifaceted advances on feature extraction, data fusion, and interpretation ability. First, GCNs extract multi-scale spatial relations by characterizing inter- and intra-interactions between different tissue regions, which are vital to understanding disease developmental mechanisms~\citep{RN1}. Further, GCNs present a strong fusion capability to handle heterogeneous cross-modality data of both imaging and non-imaging data. The cross-modality analysis is of substantial interest since it can broaden our understanding of disease mechanisms beyond the scope of single modality data. For instance, the brain networks naturally present the relationship between neurons, and the fusion with clinical records can provide auxiliary benefits for brain disease analysis~\citep{RN11}. Similarly, an integrative analysis of multi-omics profiles and imaging patterns promises to discover novel image-to-genome associations for cancer biomarker discovery~\citep{RN12,RN13}. Finally, GCNs provide a possibility of outcome interpretation by capturing the structural dynamics of complex graphs. The model outcomes can visualize both node distribution and subgraph connectivity derived from the entire graph representation. Taken together, GCNs demonstrate the potential to analyze the abundant amount of graph-level information that is crucial to advance medical imaging understanding and inform decision making in clinics.

\begin{figure}[!t]
\centering
\includegraphics[scale=.23]{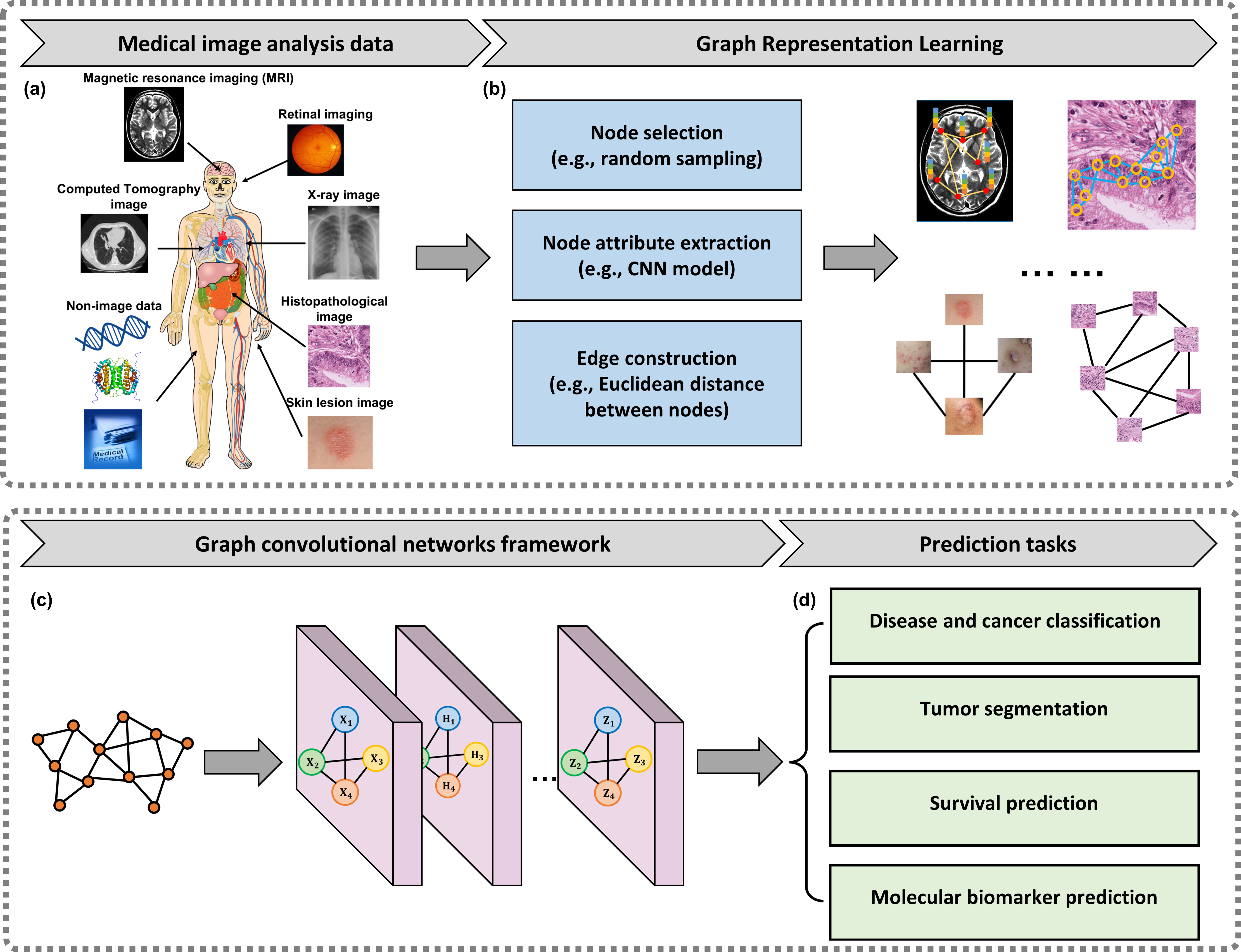}
\caption{A general pipeline for utilizing GCNs in medical image analysis.(a) Medical image analysis data. Multi-modality medical imaging and other non-image data can be jointly considered for GCN modeling and analysis. (b) Graph representation learning. The image-graph transformation pipeline includes node selection, node attribute extraction, and edge construction. For different types of medical images, we aim to design a variety of task-specific transformation strategies. (c) Graph convolutional networks framework. The input of GCNs is the constructed data-rich graphs based on image contents. The GCNs architecture contains input, hidden, and output layers to allow information extraction and inference. (d) Clinical tasks. We review a broad range of tasks with clinical relevance that incorporate disease detection, segmentation, and outcome prediction.}
\label{fig1}
\end{figure}

A general pipeline for utilizing GCNs in medical imaging is shown in Fig.~\ref{fig1}, highlighting key components of multi-modality imaging and clinical data, graph representation frameworks, and downstream clinical applications. To provide a guideline to foster cross-disciplinary research in the field of GCNs and medical imaging, the major contributions of this survey can be summarized as follows:
\begin{enumerate}
\item We outline current state-of-the-art GCNs that are widely used in medical image analysis. We summarize the key understanding of their concepts, architectures, and trade-offs of the network architecture design to advance graph and medical imaging research.
\item We highlight the convergence of graph-driven studies in radiological imaging, histopathological imaging, and other imaging modalities. We organize them in a unified taxonomy from the graph construction approaches and their downstream clinical tasks. 
\item We offer insights into the image-to-graph transformation that is vital to determine the success of GCNs, including the definition of graph components and different graph construction metrics. This review provides a key reference for researchers to explore the fast-growing synergy between graph architecture and medical imaging components.
\item Emerging opportunities and future directions are discussed in image-based GCNs and their extensions across multiple disciplines. These insights can greatly expand the scope of advancing GCNs in medical imaging and related data-driven medical studies.
\end{enumerate}

\section{Methodology of graph convolutional networks}
The architecture of graph convolutional networks (GCNs) essentially addresses the cyclic mutual dependencies with weight parameters in each network layer~\citep{RN3}. The graph convolutional layer updates graph representations by aggregating node information from their neighborhoods. Also, the edge weights and connections will be updated in specified GCNs applications. Conceptually, GCNs could broadly fall into two categories including spectral-based and spatial-based GCNs. First, the spectral-based graph convolutions are defined in the spectral domain based on the graph Fourier transformation~\citep{RN3}, which can be regarded as an analogy of the signal Fourier transform in 1-D space. Second, the spatial-based graph convolutions are defined in the spatial domain that the aggregations of node representations come from the collective information of neighboring nodes. Also, we discuss important graph pooling modules as downsampling strategies to reduce the size of graph representation~\citep{RN1}, which can critically alleviate issues of overfitting, permutation invariance, and computational complexity in the development of graph neural networks. In later sections, we define a graph as $G = (V, E)$, where $V$ is the graph node and $E$ is the edge between nodes. For graph representation learning, we use H to donate the hidden state vector of nodes.

\subsection{Spectral graph convolutional networks}
Spectral-based graph convolutional networks are derived from the field of graph signal processing, where the spectral-based convolutional operators are defined in the spectral domain~\citep{RN1}. Theoretically, a graph signal $x$ will be transformed to the spectral domain by a graph Fourier transform $\mathcal{F}$ before the convolution operation. In this way, the spectral-based graph convolutions can be computed by taking the inverse Fourier transform of the multiplication between two Fourier transformed graph signals~\citep{RN5}. Then the resulting signal is transformed back by the inverse graph Fourier transform $\mathcal{F}^{-1}$. These transformations are defined as:

\begin{equation}
\mathcal{F}(x)= U^{T}x
\end{equation}

\begin{equation}
\mathcal{F}^{-1}(x) = Ux
\end{equation}

$U$ is the matrix of eigenvectors of the normalized graph Laplacian matrix $L = I_{N}-D^{-\frac{1}{2}} AD^{-\frac{1}{2}}$, where $I_{N}$ is the normalized identity matrix, $D$ is a node degree matrix and $A$ is the adjacency matrix, which represents the connectivity between every two nodes. $L$ has the property of being real symmetric positive semidefinite. With this property, the normalized Laplacian matrix can be factorized as $L = U \Lambda U^{T},$ where $\Lambda$is a diagonal matrix of all the eigenvalues. According to the graph fourier transformation, the input graph signal $x$ with a filter $g \in R^{n}$ is defined as:

\begin{equation}
g\star x =\mathcal{F}^{-1)}(\mathcal{F}(g) \odot \mathcal{F}(x)) = U(U^{T}g\odot U^{T}x)
\end{equation}

where $\odot$ denotes the element-wise product, $U^{T}g$ is a filter in the spectral domain. If we simplify the filter by a learnable diagonal matrix $g_{\theta}= diag(U^{T}g)$, then the spectral graph convolution can be simplified as:

\begin{equation}
g_{\theta}\odot x =Ug{\theta}U^{T}x
\end{equation}

The majority of spectral-based graph convolutional networks are based on the above definitions, and the design of filter $g_{\theta}$ determines the various performance of individual approaches. Normally, the spectral-based graph convolutional network designs the convolution operation in the Fourier domain by computing the eigen-decomposition of the graph Laplacian~\citep{RN3}. They assume that the filter $g_{\theta}=\Theta_{i,j}^{(k)}$ is a set of learnable parameters and considers graph signals with multiple channels. Due to the eigen-decomposition of the Laplacian matrix, any perturbation to a graph can result in changes of eigenbasis~\citep{RN3}. The learned filters are domain dependent with a poor graph structure generalization. Also, eigen-decomposition has a high computational complexity that is unfavorable for large-scale data processing. To overcome the limitations, especially the computational complexity, Chebyshev spectral CNN (ChebNet)~\citep{RN14} used K-polynomial filters to achieve a good localization in the vertex domain by integrating the node features within the K-hop neighborhood, i.e., $g_{\theta}=\sum_{i=0}^{k}\theta_{i}T_{i}\bar{L}$, where $\bar{L} = \frac{2}{\lambda_{max}}L=I_{N}$, $\lambda_{max}$ denotes the largest eigenvalue of L. The range of the eigenvalues in $\bar{L}$ is $[-1,1]$. The Chebyshev polynomials are defined recursively as $T_{i}(x) = 2xT_{i-1}(x)-T_{i-2}(x)$ with $T_{0}(x)=1$ and $T_{1}(x)=x$. The convolution operation can be written as:

\begin{equation}
g_{\theta} \star x = \sum_{i=0}^{k}k\theta_{i}T_{i}\bar{L}x
\end{equation}

For a similar purpose of improving computational efficiency, CayleyNet~\citep{RN15} applies the Cayley polynomials that are parametric rational functions to capture narrow frequency bands. The spectral graph convolution operation is defined as:

\begin{equation}
g_{\theta} \star x =c_{\theta} x + 2Re{\sum_{j=1}^{r}c_{j}(hL-iI)^{-j)}x}
\end{equation}

Where $Re(\cdotp)$ returns the real part of a complex number, $c_{0}$is a real coefficient, $c_j$ is a complex coefficient, $i$ is the imaginary number, and $h$ is the parameter that controls the spectrum of a Cayley filter. ChebNet could be regarded as a special case of CayleyNet via the use of the Chebyshev polynomial approximation to reduce the computational complexity. 

A notable variant of ChebNet for further simplifying the computational complexity, which truncates the Chebyshev polynomial to the first-order approximation that the central node only considers its 1-hop neighboring nodes~\citep{RN16}. The approach simply filters in (5) with $i=1$ and $\lambda_{max}=2$ to alleviate the problem of overfitting: 

\begin{equation}
g_{\theta} \star x = \sum_{i=0}^{k}k\theta_{i}T_{i}\bar{L}x \approx \theta_{0}x + \theta_{1}(L - I_{N})x = \theta_{0}x - \theta_{1}D^{-\frac{1}{2}} AD^{-\frac{1}{2}}x
\end{equation}

To restrain the number of parameters and avoid overfitting, GCN further assumes that $\theta = \theta_{0} = \theta_{1}$ so that $g_{\theta} = \theta(I_{N} + D^{-\frac{1}{2}} AD^{-\frac{1}{2}})$. To solve the exploding or vanishing gradient problem in (7): $I_{N} + D^{-\frac{1}{2}} AD^{-\frac{1}{2}} \rightarrow \bar{D}^{-\frac{1}{2}}\bar{A}\bar{D}^{-\frac{1}{2}}$, with $ A = A + I_{N}$ and $\bar{D_{ij}} = \sum_{j}A_{ij}$. The propagation layer of GCN is defined as:

\begin{equation}
H = \bar{D}^{-\frac{1}{2}}\bar{A}\bar{D}^{-\frac{1}{2}}X\Theta
\end{equation}

where $X \in R^{N \times F}$is the input matrix, $\Theta \in R^{N \times F^{\prime}}$ is the parameter and $H \in R^{N \times F^{\prime}}$ is the output matrix. $F$ and $F^{\prime}$ are the dimensions of the input and the output, respectively. 

Recent research findings demonstrate the improvement of GCN’s feasibility and consistency on graph models. The adaptive graph convolution network (AGCN)~\citep{RN17} could construct and learn a residual graph Laplacian matrix for each sample in the batch through a learnable distance function that takes two nodes’ features as inputs. The residual graph Laplacian matrix leads to achieving high-level performance in public graph-structured datasets. In addition, the dual graph convolutional network (DGCN)~\citep{RN18} explores the perspective of augmenting the graph Laplacian as AGCN~\citep{RN17}. DGCN jointly considers the local consistency and global consistency on graphs through two convolutional networks. The first convolutional network is the same as (8), while the second network replaces the adjacency matrix with the positive pointwise mutual information (PPMI) matrix.

Spectral-based graph convolutional networks have a solid theoretical foundation derived from graph signals theories. Despite efforts to overcome the computation complexity, the generalization power of spectral-based GCNs is limited as opposed to the broad usage of spatial-based approaches below. Currently, the spectral-based methods train the filters on the fixed graph structure, making the trained filters unable to apply to a new graph with different structures. However, the graph structures can dramatically vary in both size and connectivity in practical applications~\citep{RN17}. The generalization power across different tasks and the high computation complexity become the major hurdles to developing spectral-based graph convolutional networks. 

\subsection{Spatial graph convolutional networks}

The spatial graph convolutional operation essentially focuses on aggregating and updating node representation by propagating node information along edges~\citep{RN3}. The aggregation strategy can directly improve the generalization power of dealing with different structured graphs by aggregating the information from neighboring nodes and updating the center node representation.

The message-passing neural network (MPNN)~\citep{RN19} represents a general framework of spatial-based GCNs~\citep{RN3}. The key forward propagation strategy of MPNN is passing the information between nodes through edges directly. As defined in the propagation function below, MPNN runs $T$ steps message-passing iterations so that the information could be propagated between nodes. Notably, GraphSAGE~\citep{RN20} is a general inductive framework which generates embeddings by sampling and aggregating features from a node’s local neighborhood. GraphSAGE leverages node feature information to efficiently generate node embeddings for previously unseen data~\citep{RN20}.

The propagation rule follows:

\begin{equation}
    h_{N(v)}^{k}  = AGGREGATE_{k}({h_{u}^{k-1}),\forall u \in N(v)})
\end{equation}

\begin{equation}
    h_{v}^{k}  =\sigma(W^{k} \cdot CONCAT(h_{v}^{k-}),h_{N(v)}^{k}
\end{equation}

Where AGGREGATE is an aggregator function that could aggregate information from node neighbors. Three types of aggregators are utilized in GraphSAGE, including mean aggregator, LSTM aggregator, and pooling aggregator. $W^{k}$is a set of weight matrices that are used to propagate information from different layers. CONCAT is the concatenated operation. Interestingly, GraphSAGE with a mean aggregator can be considered as an inductive version of GCN. To further identify the graph structures that cannot be distinguished by GraphSAGE~\citep{RN21}, Graph Isomorphism Network (GIN)~\citep{RN22} is a maximally powerful architecture to distinguish the isomorphism graph. As proved in GIN~\citep{RN22}, the injective aggregation update maps node neighborhoods to different feature vectors so that the isomorphism graph can be distinguished. To achieve the injectivity of the AGGREGATE, sum-pooling is applied in GIN. The AGGREGATE and COMBINE steps are integrated as follows:

\begin{equation}
    h_{v}^{(k)}  = MLP^{(k)}((1+ \epsilon ^{(k)}) \cdot h_{v}^{(k-1)} + \sum_{u \in N(v)}h_{u}^{(k-1)})
\end{equation}
MLP is a multi-layer perceptron that could represent the composition of functions.

The attention mechanism has been increasingly applied in spatial-based GCNs models for various sequence-based approaches\linebreak ~\citep{RN22, RN1}. Several key works are attempting to utilize attention mechanisms on graphs. Different from the design of spectral and spatial convolutional operations, the attention-based convolutional operations assign different weights for neighbors to stabilize the learning process and thus alleviate noise effects. A benefit of attention mechanisms is that they allow for dealing with variable-sized inputs, and focusing on the most relevant parts of the input to make decisions~\citep{RN22}. Graph Attention Network (GAT)~\citep{RN22} proposes a computationally efficient graph attentional layer which leverages self-attention and multi-head attention mechanisms. The GAT layer is parallelizable across all nodes in the entire graph while allowing for assigning different importance weights to different (degree) nodes in different size neighborhoods, and does not depend on knowing the entire graph structure. The coefficients computed by the attention mechanism and the propagation of GAT is formulated as:

\begin{equation}
    \alpha_{ij} = \frac{exp(LeakyReLU(\alpha^{T}[Wh_{i} ||Wh_{j}]))}{\sum_{k \in N_{i}}exp(LeakyReLU(\alpha^{T}[Wh_{i} ||Wh_{k}])}
\end{equation}

\begin{equation}
h_{i}^{\prime} = \sigma(\sum_{k \in N_{i}} \alpha_{ij} W h_{j})
\end{equation}

where  $\alpha$ and and $W$ are weight vectors, and $||$ is the concatenation operation.

Furthermore, GAT leverages multi-head attention~\citep{RN23} to stabilize the learning process of self-attention (13), which can be written as:
\begin{equation}
h_{i}^{\prime} = \prod_{k=1}^{k}(\sum_{k \in N_{i}}\alpha_{ij}Wh{j})
\end{equation}

\begin{equation}
h_{i}^{\prime} = \sigma(\frac{1}{k}\sum_{k \in N_{i}} \sum_{j \in {N_{i}}}\alpha_{ij}^{k} W_{k} h_{j})
\end{equation}

where $\alpha_{ij}^{k}$ are normalized attention coefficients computed by the k-th attention mechanism. GAT achieved significant improvement in both transductive tasks and inductive tasks, especially in the inductive task (e.g., protein-protein interaction dataset), GAT improved the micro-averaged F1 scores by 20.5\% compared to the best GraphSAGE result. 

In summary, spatial-based convolutional graph operations follow a neighborhood aggregation strategy, where we can iteratively update the representation of a node by aggregating representations of its neighbors. After k iterations of aggregation, a node’s representation captures the structural information within its k-hop network neighborhood. The rapid development of spatial-based GCNs has displayed  their computational efficiency, graph-structure flexibility, and potential generalization across tasks while compared with spectral-based GCNs~\citep{RN3}. First, spatial-based GCNs tend to be more efficient than spectral-based GCNs because they directly perform convolutions in the graph domain via node information propagation. Thus spatial-based GCNs do not have to perform eigenvector computation or handle the whole graph computation simultaneously. Second, spatial-based models are flexible to handle multi-sourced graph inputs via the convenient aggregation function~\citep{RN3}. These graph inputs can be prepared as edge inputs ~\citep{RN24, RN25, RN26, RN27, RN28}, directed graphs~\citep{RN29, RN30}, signed graphs~\citep{RN31}, and heterogeneous graphs~\citep{RN32, RN33}. Third, spatial-based models perform graph convolutions locally on each node where network weights can be efficiently generalized across different nodes and graph structures. Therefore, spatial-based models have been shown to achieve superior performance on both transductive (e.g., semi-supervised learning) and inductive (e.g., the traditional supervised learning) tasks with flexibility on graph structures. 

\subsection{Graph pooling mechanisms}
\begin{figure}[!t]
\centering
\includegraphics[scale=.15]{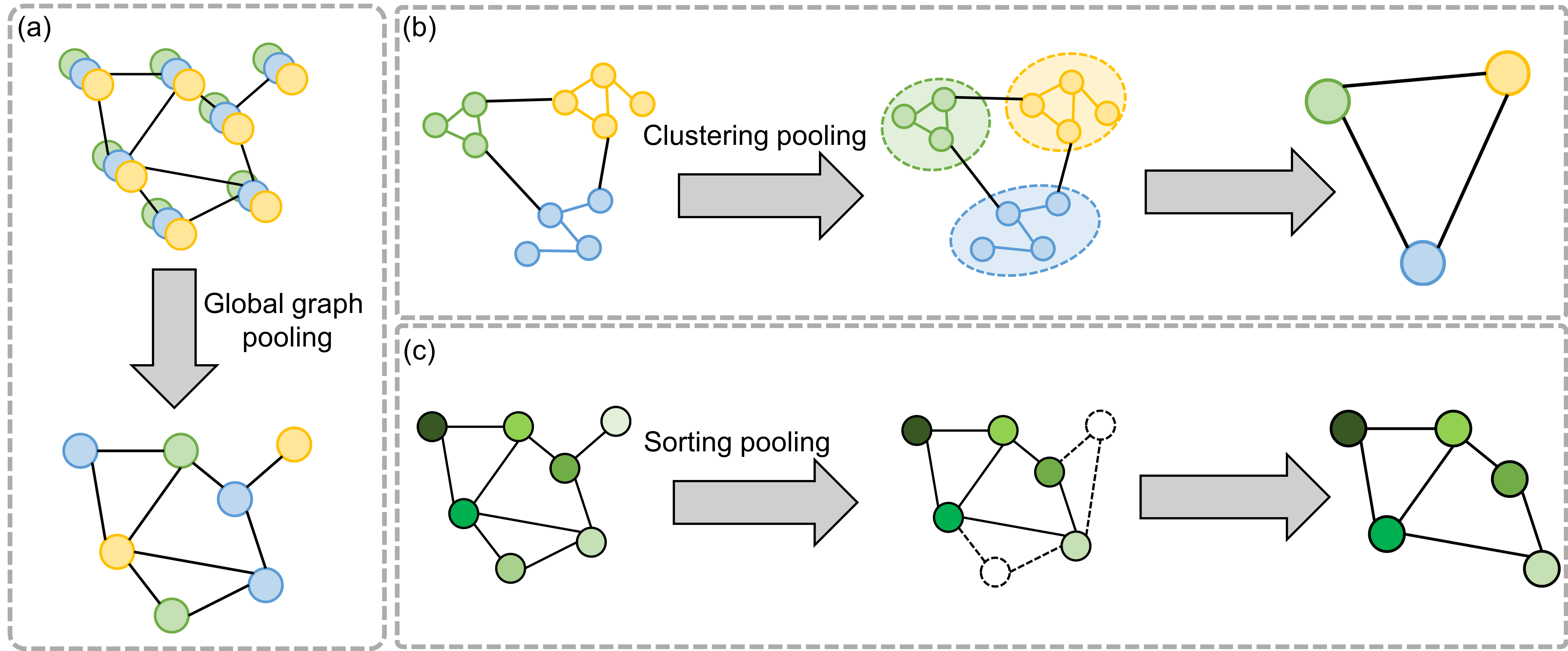}
\caption{Graph pooling mechanism. (a) Global graph pooling. The function of global graph pooling is to flatten the node representations to a graph representation. In node representation, each node will include multiple dimensions of node attributes. After utilizing the global graph pooling, the most representative feature will be selected as the node attribute in graph representation. (b) Clustering graph pooling. Clustering-based poolings offer an efficient means to find strongly-connected communities on a graph. The nodes in the same clusters are represented by a new cluster node representation. (c) Sorting graph pooling. Sorting-based pooling updates the node representation by sorting the nodes attributes or edges weights. Both (b) and (c) are hierarchical pooling operations that refine the node representation to gain model robustness and improve computation efficiency.}
\label{fig2}
\end{figure}

Graph pooling is a key strategy to address the computational challenges derived from graph convolutional operations~\citep{RN34}. Pooling operations reduce the size of a graph representation while preserving valuable structural information. Typically, graph pooling layers are located after graph convolutional layers and work as a down-sampling strategy. Graph pooling can be categorized into global and hierarchical graph poolings as shown in Fig.~\ref{fig2}. 

Global pooling operation aggregates the node representations via simple flattening procedures such as summing, averaging, or maxing the node embeddings that are widely used in graph classification tasks~\citep{RN34}. Further, a global sorting pooling~\citep{RN35} sorts the node features in a descending order based on their last feature channel and the k-largest nodes form the updated graph representation of the global sorting pooling layer. Also, global attention pooling~\citep{RN36} acts as a soft attention mechanism that decides relevant nodes to the current graph-level task. Such global-wise pooling strategies, also known as readout layers, are often used to generate graph-level representation based on the previous node representations. 

Hierarchical pooling operation is designed to refine the node representation by down-sampling strategies and overcome model overfitting. Hierarchical pooling strategies could be further categorized into two types including clustering-based and sorting-based methods. In clustering methods, spectral clustering (SC) offers an efficient means to find strongly-connected communities on a graph. SC can be used in GCNs to implement pooling operations that aggregate nodes especially belonging to the same cluster~\citep{RN37}. However, the expense of eigen-decomposition of the Laplacian and the generalization of SC strategies remain yet to be explicitly addressed. Alternatively, a graph clustering approach~\citep{RN37} formulates a continuous relaxation of the normalized min-cut problem and trains GCNs to compute cluster assignments. Spatial-based clustering strategies are proposed to achieve a higher computation efficiency compared with spectral-based clustering strategies. For example, DIFFPooling~\citep{RN38} is a differentiable graph pooling strategy that can generate hierarchical representations of graphs and can be combined with various GCNs architectures in an end-to-end fashion~\citep{RN38}. The key design of DIFFPooling is to learn a differentiable soft cluster assignment for nodes at each GCNs layer and mapping nodes to a set of clusters, which then forms the coarsened input for the next GCNs layer. In sorting-based methods, they focus on updating the node representation by sorting the nodes and edges depending on their attributes or weights. TopKPooling and SAGPooling shared a similar idea on the node sorting by their attention scores~\citep{RN39, RN40, RN41, RN42}. These poolings are designed to help select the top-kth nodes to summarize the entire graph for further feature computations. Notably, TopKPooling and SAGPooling can drop the node during model training to improve the computation efficiency and thus overcome the model overfitting. From the graph edge perspective, EdgePooling~\citep{RN43, RN44} is an inspiring example that could drop edges and merge nodes by sorting all edge scores and successively choosing the useful edges with the highest score whose two nodes have not yet been part of a contracted edge. 

\subsection{Trade-offs in the design of GCNs architectures }

To optimize the performance of graph network models, there are multiple trade-offs between the network architecture and the corresponding model performance. The ability of information collection and the strategy of effective aggregation are crucial factors for measuring the performance of GCNs models. Intuitively, a deeper architecture corresponds to a larger receptive field, which can collect more auxiliary information towards enhanced performance of GCNs. However, the performance might decrease when layers go deeper to evolve larger receptive fields in real applications~\citep{RN25}. Such performance deterioration could be attributed to the over-smoothing of node representation with an increased architecture depth. In other words, the repeated and mixed message aggregation can lead to node representations of inter-classes indistinguishable~\citep{RN45}. It is commonly seen that the over-smoothing issue always occurs in the nodes with a dense connection with other nodes (e.g., the core of the graph) that could rapidly aggregate information in the entire graph. In contrast, the node in the tree part (e.g., leaves of the tree) could only include a very small fraction of information of all nodes with a small number of GCNs layers. To improve the GCNs model performance, it is necessary to overcome the graph over-smoothing phenomena and achieve informative node representation. For example, the study~\citep{RN46} implemented a co-training and self-training scheme with a smoothness regularizer term and adaptive edge optimization~\citep{RN45} to alleviate the over-smoothing problem. Co-training a GCN with the random walk model can explore the global graph topology. Further, self-training a GCN could exploit feature extraction capability to overcome its localized limitation. Informative node representation via the jumping knowledge network (JK-Net)~\citep{RN47} tends to demonstrate compelling performance on graph computing efficiency and alleviate overfitting. Notably, the idea of layer-aggregation across layers helps select the most informative nodes and reduce the overfitting issue, and the LSTM-attention could further identify the useful neighborhood ranges. Inspired by the architecture of JK-Net, Deep adaptive graph neural network (DAGCNs)~\citep{RN25} developed an adaptive score calculation scheme for each layer, which could balance the information from both local and global neighborhoods for each node. Both JK-Net and DAGCNs aim to find a trade-off between accuracy performance and the size of receptive fields by adaptively adjusting the information from local and global neighborhoods. For the design of network architecture, we expect additional efforts to overcome the over-fitting issues while keeping a flexible architecture to explore more meaningful information in the context of disease detection and diagnosis.

\section{Development of GCNs in medical imaging}

\subsection{Radiological image analysis}

Over the past decades, multi-modality radiological images have been routinely utilized in abnormality segmentation~\citep{RN48, RN49}, detection~\citep{RN50, RN51}, and patient outcome classification~\citep{RN52, RN53}. In this section, we discuss the growing body of GCNs studies applied to radiological analysis~\citep{RN54, RN55, RN56}, including magnetic resonance imaging (MRI), Computed Tomography (CT), and X-ray imaging. The combination of GCNs and radiological imaging promises to reflect the interaction among tissue regions and provide an intuitive means to fuse the morphological and topological-structured features among key image regions to advance modeling, interpretation, and outcome prediction. We here discuss the representative neuroimaging research and other related studies to highlight the usefulness of GCNs across different radiological imaging modalities and clinical tasks.

\subsubsection{Neuroimaging}
\begin{figure}[!t]
\centering
\includegraphics[scale=.12]{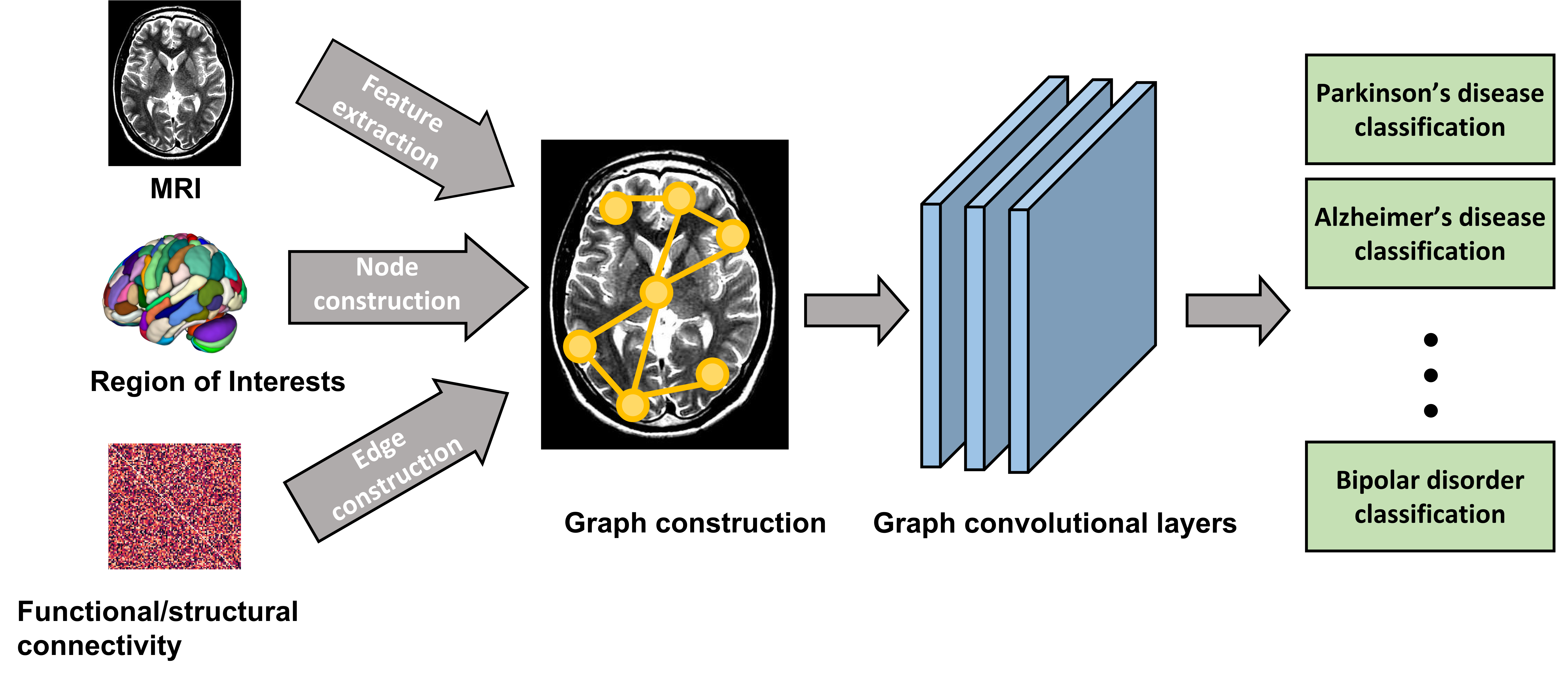}
\caption{The framework of developing GCNs in neuro-imaging analysis. Multi-modality MRIs are firstly converted into graph structure which is determined by the region of interest in terms of real human brain signals or fiber connectivity (e.g., node and edge definitions). Through graph-level model development and inference, we highlight numerous image-based analysis and diagnosis of diseases in neurology.}
\label{fig3}
\end{figure}

In neuroimaging, multi-modality MRI is a useful diagnostic technique by providing high-quality three-dimensional (3D) images of brain structures with detailed structural information~\citep{RN57}. Conceptually, multi-modality MRI data can be categorized into functional MRI (fMRI), structural MRI (sMRI), and diffusion MRI (DMRI). The fMRI measures brain activity and detects the changes in blood oxygenation and blood flow in response to neural activity~\citep{RN58}. The sMRI translates the local differences in water content into different shades of gray that serve to outline the shapes and sizes of the brain’s various subregions~\citep{RN59}. The DMRI is a magnetic resonance imaging technique in which the contrast mechanism is determined by the microscopic mobility of water molecules~\citep{RN60}. All these imaging modalities provide vital diagnostic support for neurological disease analysis because they can capture anatomical, structural, and diagnosis-informative features in neurology. Therefore, the overarching goal is to develop useful graph network models to define, explore, and interpret interactions of brain neurons and tissues. The detailed process of utilizing GCNs in the neuro-imaging analysis is illustrated in Fig.~\ref{fig3}. 

To analyze the complex brain region connectivity and interaction, a brain graph representation can intuitively portray human brain organization, neurological disorders, and associated clinical diagnosis. Conventionally, the human brain could be modeled into a brain biological network containing nodes (e.g., region of interests) and edges among brain network nodes. The edges could be determined by brain signals or the real fiber connection. Yet these biologically-defined networks are often unable to faithfully capture neurological disorders and outcomes of patients~\citep{RN61}. To overcome this challenge, it is encouraged to leverage informative image-based features  to considerably enrich graph node attributes. Comprehensive graph representation can integrate multiple types of information (e.g., image features, human brain signals, and clinical data) to greatly expand the knowledge base of brain dynamics and potentially provide auxiliary clinical diagnosis assistance. The use of GCNs here can be helpful to augment the architecture of human brain networks and has achieved remarkable progress in explaining the functional abnormality from the network mechanism~\citep{RN62}. In particular, GCNs are able to consider the functional or structural relations among brain regions together with image-based features that are beyond the scope of the conventional CNN-based methods~\citep{RN63,RN64,RN65}. The CNN-based model is merely viewed as a feature extractor for disease representation without consideration of structure information of the brain. For example, the deep 3-D convolutional neural network architecture was not unable to capture underlying structure information for Alzheimer's disease classification using brain MRI scans~\citep{RN61}. By contrast, the convergence of GCNs methods and MRI provide an alternative means to characterize the architecture of human brain networks and has achieved outstanding progress in brain abnormality explanation~\citep{RN62}. 

The graph representations can be divided into functional and structural brain connectivity graphs based on the definitions of the graph components. First, graph nodes are regions of interest (ROI) as defined in MRI. ROI definition is commonly done through the anatomical parcellation of the Montreal neurological institute (MNI) using sMRI and fMRI data~\citep{RN66,RN67,RN68}. Second, graph edges are determined by the physical connectivity (e.g., the fiber tracts) of nodes in structural brain networks while calculated from the signal series analysis in functional brain networks.  We therefore discuss insights of functional and structural brain connectivity graph developments below.

The human brain functional connectivity denotes the functional relations between specific human brain areas and functional brain graphs can represent estimates of interactions among time series of neuronal activity~\citep{RN62}. In functional brain networks, the nodes are defined as brain parcellation ROIs while the node attributes could be hand-crafted features or correlation measurements between nodes. The edges are created through the node correlations between different regions. For example, the GCNs framework achieved high-level performance in classifying Autism spectrum disorders (ASD) and healthy controls (HC) using task-functional magnetic resonance imaging (task-fMRI) through the appropriate ROI definition~\citep{RN69}. Their model consists of a message-passing neural network (MPNN)~\citep{RN19} as convolutional layers that is invariant to graph symmetries~\citep{RN69}. Furthermore, Top-k poolings~\citep{RN39} is able to downsample the node to achieve a higher computation efficiency while preserving a meaningful graph delineation. For a similar ASD and HC classification, another study~\citep{RN70} used GAT~\citep{RN22} as the convolutional layer incorporated with Top-k~\citep{RN39, RN40} and SAGE~\citep{RN42} pooling for achieving node importance scores. They also introduced two distance losses to enhance the distinguish among the nodes. Also, a group-level consistency loss is added to force the node importance scores to be similar for different input instances in the first pooling layer. Inspired by metric learning, a siamese graph convolutional neural network (s-GCN) is proposed for the ASD and HC classification purpose~\citep{RN54}, where samples were collected from Autism Brain Imaging Data Exchange (ABIDE)~\citep{RN71} database and UK Biobank~\citep{RN72}. The graph metric learning method essentially utilized  GCN layer~\citep{RN16} in a siamese network~\citep{RN73}. Two types of graph construction methods are designed as the input of the model, such as spatial and functional graphs which determine nodes by ROIs, meanwhile, the KNN algorithm is utilized for both spatial and functional graph construction. The inputs for s-GCN are two same structure spatial or functional graphs with different signals (e.g., rows or columns of functional matrix).   Notably, the spatio-temporal graph could be used to analyze the functional dependency between different brain regions and the information in the temporal dynamics of brain activity simultaneously. The spatio-temporal graph convolutional network is utilized to analyze the Blood-oxygen-Level-dependent (BOLD) signal of resting-state fMRI (rs-fMRI) for human age and sex prediction~\citep{RN74}. Also, studies~\citep{RN75, RN76} analyze the BOLD signal of fMRI for accurate detection of cognitive state changes of the human brain by presenting a dynamic graph learning approach to generate an ensemble of subject-specific dynamic graph embeddings. The brain networks are able to disentangle cognitive events more accurately than using the raw BOLD signals. The functional graph is meaningful to reflect the average functional connection strength between pairs of brain regions within a population. Generally, Pearson’s correlation is a useful strategy to construct functional connectivity matrix and define the node attributes. 

The human brain's structural connectivity in vivo can be captured by structural and diffusion MRI~\citep{RN77, RN78}, and structural brain graphs could represent anatomical wiring diagrams~\citep{RN62}. Similar to the definition of nodes in functional connectivity networks, the nodes in structural connectivity networks are defined as a region of interests (ROIs), which are parceled from the brain based on structural and diffusion MRI.. Clinically, the structural brain connectivity represents the structural associations of altered neuronal elements, including both the morphometric alternation and accurate anatomical connectivity as seen in imaging. In the complex brain networks, structural brain connectivity assesses to white amount projections bond cortical and subcortical regions~\citep{RN79}. The edges indicate the actual neural fiber connections between different brain regions. For example, a stack architecture design of combining a heterogeneous GCN model~\citep{RN55} with an efficient adaptive pooling scheme~\citep{RN38} is able to predict the clinical score of Parkinson’s disease (PD) and HC using diffusion-weighted MRI (DWI) on Parkinson Progression Marker Initiative (PPMI) ~\citep{RN80}. To construct the graph structure from DWI, nodes are defined by the ROIs in the brain while three whole-brain probabilistic tractography algorithms are able to determine different brain structural. The node attributes corresponding to rows in the human brain network are defined as features. Another novel framework is developed to explore graph structure in the q-space by representing DMRI data and utilizing graph convolutional neural networks to estimate tissue microstructure~\citep{RN81}. This approach is capable of not only reducing the data acquisition time but also accelerating the estimation procedure of tissue microstructure. The nodes of the weighted graphs are sets of points on a manifold. Also, the adjacency weights are defined between two nodes using Gaussian kernels, accounting for differences in gradient directions and diffusion weightings. The q-space signal measurements are represented by using the constructed graph that encodes the geometric structure of q-space sampling points. A residual ChebNet~\citep{RN14} can learn the mapping between sparsely sampled q-space data and high-quality estimates of microstructure indices. 

Beyond single-modality MRI analysis, multi-modality image data analysis emerged as active research areas for GCNs modeling. Multi-modality MRI data analysis is able to deepen our understanding of disease diagnosis from different data aspects. In neuroimaging, the structural connectivity in sMRI reflects the anatomical pathways of white matter tracts connecting different regions, whereas the functional connectivity in fMRI encodes the correlation between the activity of brain regions. A unique advantage of multi-modality MRI data analysis is that they have incorporated complementary information from different modalities simultaneously. The multimodal data fusion can be implemented by two types of strategy: (1) constructing the original graphs directly using the partial information from functional and structural brain connectivity; (2) constructing the original functional and structural graphs separately and updating the graph representations by computing and fusing the two-side information. For the first fusion strategy, the study~\citep{RN82} introduced an edge-weighted graph attention network (EGAT)~\citep{RN22} with a diffPooling~\citep{RN38} to classify Bipolar disorder (BP) and HC from sMRI and fMRI in cerebral cortex analysis. Also, the framework of Siamese community-preserving graph convolutional network (SCP-GCN)~\citep{RN83} is able to learn the structural and functional joint embedding of brain networks on two public datasets (i.e., Bipolar and HIV dataset~\citep{RN83}) for brain disease classification. Especially, siamese architecture can exploit pairwise similarity learning of brain networks to guide the learning process to alleviate the data scarcity problem~\citep{RN83}. Ninety cerebral regions are selected as nodes for both structural (e.g., Diffusion Tensor Image (DTI)) and functional (e.g., fMRI) networks, and the node attribute is determined by the functional connectivity between nodes corresponding to fMRI. The edge connectivity is determined by the DTI via a series of preprocessing (distortion correction, noise filtering, repetitive sampling from the distributions of principal diffusion directions for each voxel). To preserve the community property of brain networks, the design of a community loss presents its usefulness to minimize the intra-community loss and maximize the intercommunity loss. For the second fusion strategy, the study~\citep{RN84} fused information from multimodal brain networks on rs-fMRI and dMRI for age prediction. After constructing the original functional and structural brain networks separately, the study reconstructed the positive and negative connections in the functional networks depending on structural networks' information. They utilized a multi-stage graph convolutional layer, motivated by GAT~\citep{RN22} and ResGCN~\citep{RN85}, for structural network edge feature update and class classification. Then, the edge feature and class are utilized to update the functional networks for age prediction. Also, for liver lesion segmentation, a mutual information-based graph co-attention module~\citep{RN86, RN87} is proposed by extracting modality-specific features from T1-weighted images (T1WI) and establishing the regional correspondence from T2-weighted images (T2WI) simultaneously. In this study, they constructed two separate graphs for either T1WI or T2W2; meanwhile, they used GCN~\citep{RN16} to propagate representations of all nodes. The mutual information-based graph co-attention module updates the T1WI-based node representation by selectively accumulating information from node features from T2WI-based node representation. The fused node representation is re-projected and added to the original feature for the final segmentation. 

Furthermore, GCNs extend to allow the multi-modality integration between MRI and non-imaging data for analyzing complex disease patterns. For example, an Edge-Variational GCN (EV-GCN)~\citep{RN11} could automatically integrate imaging data (e.g. fMRI data) with non-imaging data (e.g. age, gender and diagnostic words) in populations for uncertainty-aware disease prediction. They constructed weighted graphs via an edge-variational population graph modeling strategy. In the weighted graphs, the graph nodes are ROIs and the node attributes are features extracted from histology and fMRI images. It is particularly notable that the weight of the edge is achieved by a learnable function of their non-imaging measurements. The proposed Monte-Carlo edge dropout randomly drops a fraction of edges in the constructed graphs to reduce overfitting and increase the graph sparsity. In addition, two similar studies~\citep{RN88, RN89} constructed the sparse graph by combining the information from the functional connectivity (e.g., rs-fMRI),  structural connectivity (e.g., DTI), and demographic records (e.g., gender and age) for mild cognitive impairment detection and classification. In these studies, they constructed functional and structural brain networks for each subject (e.g., image). Then, they defined each subject as graph node. For the first study~\citep{RN88}, they concatenated the feature from functional and structural connectivity as vertices features. Also, they calculate the feature and phenotypic information similarity to constructed graph edges. Further, they utilized GCN~\citep{RN16} incorporated with the random walk algorithm to enhance the detection performance.  For the second study~\citep{RN89}, it constructed graphs for the functional and structural connectivity separately. Beyond the similarity evaluation of subjects (e.g., vertices) feature and non-image phenotypic information, this study also determined the edge by connecting nodes belong to the same receptive field class directly. Further, they used the constructed graph to pretrain GCN to update graphs and GCN for the final disease deterioration prediction.

\subsubsection{X-ray and CT imaging}
Extensive studies have also utilized GCNs in X-ray and Computed Tomography (CT) images for disease analysis~\citep{RN56, RN90, RN91}. Different from MRI data, CT images are able to reflect the vessel skeleton information that could assist a variety of clinical tasks. For example, chest CT scans can assist with arteries-veins separations that are of great clinical relevance for chest abnormality detection~\citep{RN90}. The graph was constructed of the voxels on the skeletons resulting in a vertex set and their connections in an adjacency matrix. The skeletons are extracted from chest CT scans by vessel segmentation and skeletonization. In this study~\citep{RN90}, GCN layers can extract and learn connectivity information. The one-degree (direct) neighbors were considered and the vertices attributes were extracted by CNN model to consider the local image information. In addition, chest CT scans together with GCN is able to assist airway semantic segmentation, which refers to the segmentation of airway from background and dividing it into anatomical segments for lung lobe analysis~\citep{RN92}.  Also, a prototype-based GCN framework~\citep{RN93} provided a means for airway anomaly detection to aid in lung disease diagnosis. The GCN layers calculated the initial anomaly score for every node, while the prototype-based detection algorithm computed the entire graph's anomaly score. Another study~\citep{RN94} utilized radiotherapy CT (RTCT) and PET, which is registered to the RTCT for lymph node gross tumor volume (GTV\_LN) detection. In this study, the 3D CNN extracted instance-wise visual features while the GNN model analyzed the inter-LN relationship. The feature fusion of CNN and GNN boosted the GTV\_LN detection performance. Furthermore, the study~\citep{RN95} utilized GCN~\citep{RN16} on cone-beam computed tomography (CBCT) images for craniomaxillofacial (CMF) landmark localization, which is important for designing treatment plans of reconstructive surgery. They utilized an attention feature extraction network for localizing landmarks and generating attention features for the graph construction. In addition, the study~\citep{RN56} proposed an end-to-end hybrid network to train a CNN and GAT network to leverage both advanced feature learning and inter-class feature representations on Chest-Xray 14 dataset~\citep{RN96}. To utilize the image sequencing information, they determine each image from the same patient as a vertice of a graph and the extracted features are the attributes of vertices. Furthermore, they leverage non-imaging meta-data, such as clinical information, to construct edges between the vertices. After constructing the graph and updating the graph representation with GAT, they combine the CNN extracted features with graph representation by skip-connectivity to achieve hybrid representation. The motivation of generating hybrid representation is to improve the distinction between samples. Furthermore, CT images and non-image clinical information could be analyzed jointly for Lymph node metastasis (LNM) prediction. The study~\citep{RN97} proposed a co-graph convolutional layer consisting of Con-GAT~\citep{RN22} and Corr-GAT~\citep{RN98} layers to achieve the node’s new representation by weighted averaging its neighboring nodes and measuring the weight score by feature difference-based correlation. Due to the pandemic of COVID-19, GCNs have also been utilized in disease detection. GraphCovidNet ~\citep{RN91} utilized GIN for COVID-19 detection on both CT and X-ray images. The graph is used to depict the outline of an object (e.g., organ) in the image. First, they applied edge detection to determine the edge outline. Then, the graph nodes are defined by the pixel having a grayscale intensity value greater than or equal to 128, which implies nodes reside only on the prominent edges of the edge image. The node attribute consists of the grayscale intensity of the corresponding pixel. An edge exists between the two nodes which represent neighboring pixels in the original image. For example, the GCN-based model~\citep{RN99} extractes node information hierarchically towards both diagnosis and prognosis for COVID-19 patients. Their distance-aware pooling, including graph-based clustering and feature pooling, is able to aggregate node information on the dense graph effectively. Also, the proposed model could coarsely localize the most informative slices for CT scans to provide the interpretability for better clinical decision-making.

Table.~\ref{table1} summarizes a variety of graph construction methods and GCNs application in radiologic image analysis. Compared to conventional methods, GCNs methods for the analysis of brain networks have the possibility of combining image-based features with the conventional brain networks.

\begin{longtable}{|p{2.7cm}|p{1.6cm}|p{1.6cm}|p{10.5cm}|}
\caption{\label{table1}Summary of GCNs in radiologic image analysis}\\

\hline
 Method & Category & Input & Graph Construction\\
\hline

\endhead 
\multicolumn{4}{r}{\footnotesize Continue on the next page}
\endfoot
\endlastfoot

\multirow{2}{*}{\cite{RN69}} & \multirow{2}{*}{\makecell{Single\\modality}} & \multirow{2}{*}{Task-fMRI} & Nodes are ROIs and node attributes are hand-craft features.\\
\cline{4-4}
 & & & Edges are determined by region-to-region correlations and edge attributes are the values of Pearson correlation and partial correlation among nodes.\\
\hline
\multirow{2}{*}{\cite{RN54}} & \multirow{2}{*}{\makecell{Single\\modality}} & \multirow{2}{*}{fMRI} & Nodes are ROIs and node attributes are rows/column of the functional connectivity matrix\\
\cline{4-4}
 & & & Spatial graph: a KNN graph based on spatial coordinates of the ROI.\par Functional graph: a KNN graph based on correlation distance between all ROI pairs.\\
\hline
\multirow{2}{*}{\cite{RN70}} & \multirow{2}{*}{\makecell{Single\\modality}} & \multirow{2}{*}{fMRI} & Nodes are ROIs and node attributes are determined by Pearson correlation coefficient among nodes.\\
\cline{4-4}
 & & & Edge is determined by Edge-set and edge attributes are determined by Partial correlation coefficient among nodes.\\
\hline
\multirow{2}{*}{\cite{RN74}} & \multirow{2}{*}{\makecell{Single\\modality}} & \multirow{2}{*}{rs-fMRI} & Spatial graph: nodes are ROIs at the same time point and node attributes are determined by the average BOLD signal of the ROI i at time point t.
\par
Temporal graph: the same ROI at the different time points and node attributes are determined by the functional affinity between ROIs. The functional affinity is calculated by the magnitude of correlation between the concatenated average BOLD time series.
\\
\cline{4-4}
 & & & Spatio graph: connect all nodes at the same time point.\par
Temporal graph: connect the corresponding node to the node of the same ROI at the proceeding time point. \\
\hline
\multirow{2}{*}{\makecell{\cite{RN75}\\ \cite{RN76}}} & \multirow{2}{*}{\makecell{Single\\modality}} & \multirow{2}{*}{fMRI} & Nodes are defined by ROIs (e.g., Brain regions), and the node attributes are hand-craft features.\\
\cline{4-4}
 & & & Edges are determined by Edge-set and edge attributes are the statistical correlations of BOLD signals among nodes\\
\hline
\newpage
\multirow{2}{*}{\cite{RN55}} & \multirow{2}{*}{\makecell{Single\\modality}} & \multirow{2}{*}{DWI} & Node are ROIs, and the node attributes are the rows/columns of the connectivity matrix.\\
\cline{4-4}
 & & & Edges are determined by the whole-brain probabilistic tractography algorithm\\
\hline
\multirow{2}{*}{\cite{RN81}} & \multirow{2}{*}{\makecell{Single\\modality}} & \multirow{2}{*}{DMRI} & Nodes are the points on the manifold.\\
\cline{4-4}
 & & & Edges are constructed between two nodes when edge attributes are larger than 0, and edge attributes are determined by the differences in gradient directions and diffusion weighting calculated by two Gaussian kernels in q-space.\\
\hline
\multirow{2}{*}{\cite{RN90}} & \multirow{2}{*}{\makecell{Single\\modality}} & \multirow{2}{*}{CT images} & Nodes are the connectivity of voxels, and node attributes are extracted by CNN model.\\
\cline{4-4}
 & & & Edges are constructed between the voxels on the skeletons.\\
\hline
\multirow{2}{*}{\cite{RN92}} & \multirow{2}{*}{\makecell{Single\\modality}} & \multirow{2}{*}{CT images} & Airway nodes: nodes are the segmented airway regions, and node attributes are extracted by CNN model. \par Landmark nodes: nodes are the detected landmark positions, and node attributes are extracted by CNN model.
\\
\cline{4-4}
 & & & 1.Internal graph: the connectivity between nodes and their KNN neighbors in Euclidean space. \par 2.External graph: the airway structural prior information to connect two types of nodes. 
\\
\hline
\multirow{2}{*}{\cite{RN93}} & \multirow{2}{*}{\makecell{Single\\modality}} & \multirow{2}{*}{CT images} & Nodes are the segmented airway regions, and node attributes are extracted by CNN model and the anatomical airway structure.
\\
\cline{4-4}
 & & & The edges are determined by the airway structure.
\\
\hline
\multirow{2}{*}{\cite{RN94}} & \multirow{2}{*}{\makecell{Single\\modality}} & \multirow{2}{*}{\makecell{RTCT and\\PET images}} & Nodes are the $GTV_{LN}$ candidates, and node attributes are extracted by CNN model.
\\
\cline{4-4}
 & & & The edges are determined by calculating the Euclidean distance among the boundary voxels of the tumor mask.
\\
\hline
\multirow{2}{*}{\cite{RN95}} & \multirow{2}{*}{\makecell{Single\\modality}} & \multirow{2}{*}{\makecell{CBCT\\images}} & Nodes are the landmarks, and node attributes are extracted by CNN model.
\\
\cline{4-4}
 & & & The edge between landmarks is determined by whether the two landmarks are in the same anatomical regions .
\\
\hline
\multirow{2}{*}{\cite{RN99}} & \multirow{2}{*}{\makecell{Single\\modality}} & \multirow{2}{*}{CT images} & Nodes are the slices of CT image, and node attributes are extracted by CNN model and wavelet decomposition.
\\
\cline{4-4}
 & & & The graph is densely connected graph, and edge attributes are calculated by cosine similarity among node features.
\\
\hline
\multirow{2}{*}{\cite{RN82}} & \multirow{2}{*}{\makecell{multi\\modality}} & \multirow{2}{*}{\makecell{fMRI \\and sMRI}} & Nodes are ROIs, and node attributes are seven anatomical features and four functional connectivity statistic features.
\\
\cline{4-4}
 & & & The graph is densely connected graph, and edge attributes are calculated by the Pearson correlation-induced similarity..
\\
\hline
\multirow{2}{*}{\cite{RN83}} & \multirow{2}{*}{\makecell{multi\\modality}} & \multirow{2}{*}{\makecell{fMRI \\and DTI}} & Nodes are ROIs, and node attributes are rows/column of connectivity matrix.
\\
\cline{4-4}
 & & & The edges are determined by region-to-region correlations.
\\
\hline
\multirow{2}{*}{\cite{RN84}} & \multirow{2}{*}{\makecell{multi\\modality}} & \multirow{2}{*}{\makecell{rs-fMRI \\and dMRI}} & Nodes are ROIs.
\\
\cline{4-4}
 & & & The edges are determined by edge-set. For functional brain network, the edge attributes are the correlation of fMRI signals between nodes. For structural brain network, the edge attributes are the probability of fiber tractography between nodes. 
\\
\hline
\multirow{2}{*}{\makecell{\cite{RN86}\\\cite{RN87}}} & \multirow{2}{*}{\makecell{multi\\modality}} & \multirow{2}{*}{\makecell{T1WI \\and T2WI}} & Nodes are a group of features in the partial region of the original regular grid coordinates, and node attributes are extracted by CNN model.
\\
\cline{4-4}
 & & & The graph is a fully-connected graph.
\\
\hline
\newpage
\multirow{2}{*}{\cite{RN11}}&\multirow{2}{*}{\makecell{multi\\modality}} & \multirow{2}{*}{\makecell{fMRI and\\clinical data}} & Nodes are images (e.g., subjects), and node attributes are the concatenated local weighted clustering coefficient features from functional and structural connectivity.
\\
\cline{4-4}
 & & & The edges are determined by calculating the similarity between the node features and incorporates the phenotypic information.
\\
\hline

\multirow{2}{*}{\cite{RN88}} & \multirow{2}{*}{\makecell{multi\\modality}} & \multirow{2}{*}{\makecell{rs-fMRI,\\DTI, gender\\ and age}} & Nodes are images (e.g., subjects), and node attributes are the the concatenated local weighted clustering coefficient features from functional and structural connectivity.
\\
\cline{4-4}
 & & & The edges are determined by the similarity between the node features and incorporates the phenotypic information.
\\
\hline
\multirow{2}{*}{\cite{RN89}} & \multirow{2}{*}{\makecell{multi\\modality}} & \multirow{2}{*}{\makecell{rs-fMRI,\\DTI, gender\\ and\\acquisition\\equipment}} & Nodes are images (e.g., subjects), and node attributes are the recursive feature elimination extracted features from functional and structural connectivity.
\\
\cline{4-4}
 & & & The edges are determined by:\par
 1. The similarity between the node features and incorporates the phenotypic information.\par
2. Connected nodes that belong to the same receptive field.
\\
\hline

\multirow{2}{*}{\cite{RN56}} & \multirow{2}{*}{\makecell{multi\\modality}} & \multirow{2}{*}{\makecell{X-ray\\images\\and\\meta-data}} & Nodes are images (e.g., patients), and node attributes are extracted by CNN model.
\\
\cline{4-4}
 & & & The edges are determined by non-image meta-data.
\newline
\\
\hline
\multirow{2}{*}{\cite{RN97}} & \multirow{2}{*}{\makecell{multi\\modality}} & \multirow{2}{*}{\makecell{CT images\\and\\non-image\\data}} & Nodes are ROIs, and node attributes are the concatenation feature consists of CNN extracted features and non-imaging clinical information.
\newline
\\
\cline{4-4}
 & & & The graph is fully connected graph.
\\
\hline
\multirow{2}{*}{\cite{RN91}} & \multirow{2}{*}{\makecell{multi\\modality}} & \multirow{2}{*}{\makecell{CT and\\X-ray\\images}} & Nodes are pixels, and node attributes are grayscale intensity of the pixel.
\\
\cline{4-4}
 & & & The edges are determined by the neighborhood relationship between pixels.
\newline
\\
\hline
\end{longtable}

\subsection{Histopathological image analysis }
The growth of digitalized histopathological images presents a valuable resource to support rapid and accurate clinical decision making. The high-resolution whole slide image (WSI) contains rich tissue characteristics including patterns of cell nuclei, glands, and lymphocytes~\citep{RN100, RN101}. Extensive pathological characteristics of tissue and cell interactions can be evidently observed that are not available in other clinical image data. For instance, lymphocytic infiltration of cancer status can be deduced only from histopathology imagery~\citep{RN102}. These pathological patterns can be used to build the biological graph networks that can inform disease status and thus discern predictive imaging biomarkers. Overall, we recognize that GCNs analysis is uniquely positioned to address key issues of histopathological applications, including data annotation, tissue connections, global-local information diagnostic fusion, and model prediction performance in challenging settings. 

Developments of GCNs have brought remarkable advances into computational histopathology including label efficiency and multi-scale context representation. First, graph structure provides a reasonable choice to represent the entire slide in terms of tissue content connectivity. Such entire-slide graph representation can avoid fine-grained patch-wise label annotation. Since we know that patch-level labeling is highly time-intensive, even impossible, to include all ranges of tumor patches annotated by human experts. Second, graph structural representation can capture multi-scale contexts considering both global and local image-wise features towards enhanced prediction of disease outcomes. Third, graph structural representation builds upon the interaction among spatially-separated tiles that enables a more flexible and comprehensive receptive field. Such advances are analogous to the workflow of human experts that we consider tumor environment, tissue contents, and their interactions, rather than single tumor tiles, to diagnose tissue status of patients.

\begin{figure}[!t]
\centering
\includegraphics[scale=.22]{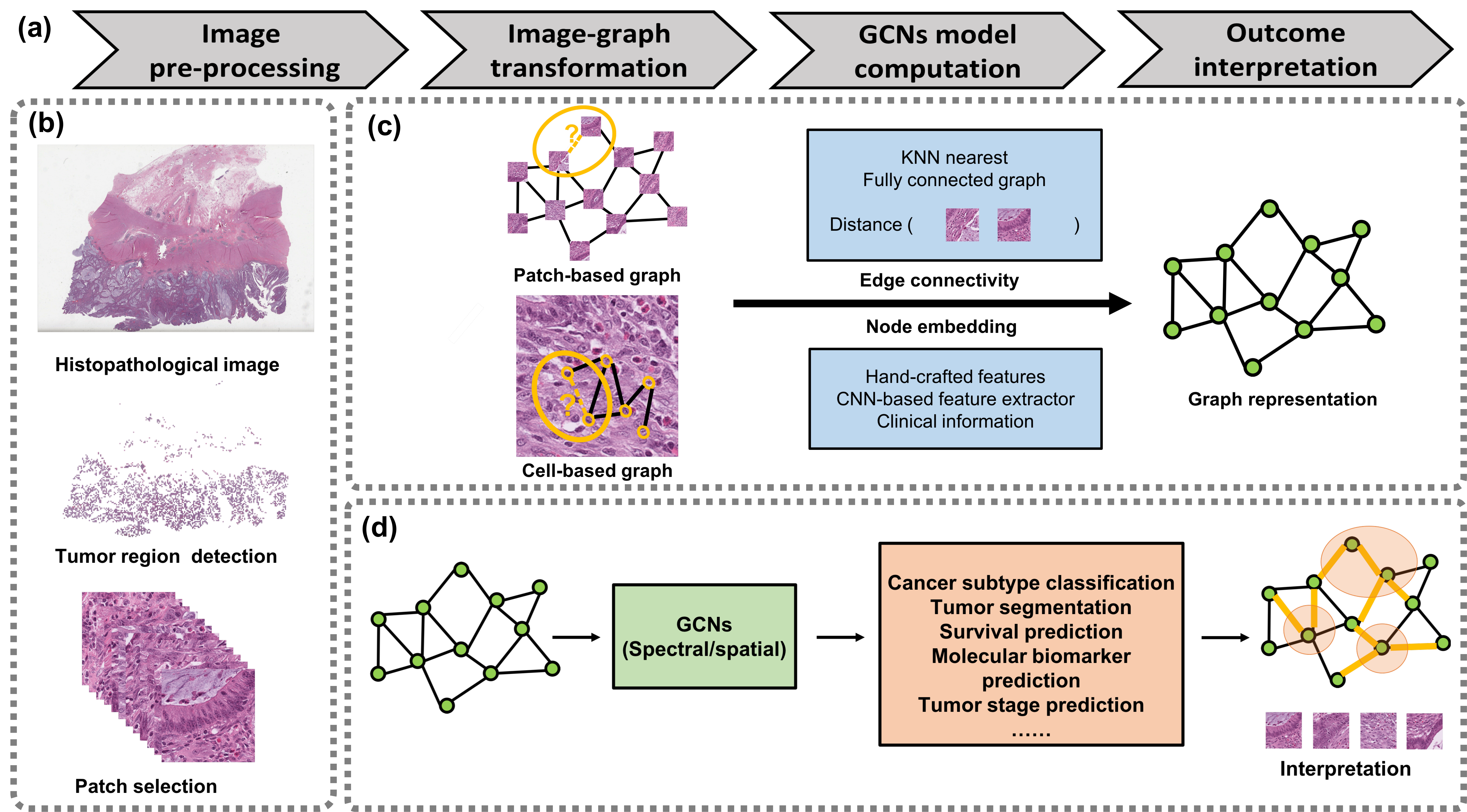}
\caption{The illustration of the computational framework of GCNs in histopathological imagery. (a) Overall steps of image-based graph convolutional network framework. (b) Image preprocessing. The high-resolution images are normally split into manageable small-sized tumor tiles. We primarily focus on tumor tile processing and analysis for GCN development. (c) Image-based graph transformation. The transformation between image and graph-structured data is vital according to different tasks. Both patch-based and cell-based graphs can be established for downstream tasks. (d) GCNs computation and outcome interpretation. The inputs of GCNs are the constructed image-based graphs. The outcome interpretation of GCNs includes both node- and edge-wise findings to enable a multi-dimensional interpretation of outcomes.}
\label{fig4}
\end{figure}

Because the high-resolution histopathological image does not present a natural form of graph structure, efficient graph representation becomes a vital factor for model development and optimization. Current graph construction in histopathology can be broadly categorized into patch-based and cell-based methods. First, patch-based graph construction methods aim to enable information extraction by considering the entire micro-environment (e.g., the cells and tissues), where comprehensive tissue micro-environment and cell dynamics can be captured. In these patch-based methods, graph nodes are defined as the selected patches determined by ROIs in the histopathological image. The associated node attributes can be extracted by standard feature extractors (e.g., ResNet18 or VGG16). Graph edges are defined as the connectivity between nodes, which is determined by the feature or coordinate distance between two nodes. A smaller distance means a higher probability of connectivity. The connectivity between nodes could determine an adjacency matrix to represent the entire topological structure of the graph. Although the definition of primitive graph components (e.g., node and edge) are conceptually similar, most patch-based graph construction methods have different settings for node attributes and edge construction. As opposed to the patch-based graphs, cell-based graph methods emphasize the possible biological significance derived from histopathology. Cell-based graph construction methods aim to model the relationship between different cells and the micro-environment (e.g., tissues or vessels) utilizing graph-based features~\citep{RN103}. In a cell graph, the detected and segmented nuclei or cell clusters are considered as nodes. The node attribute is defined as the combination of image-wised features, such as features extracted by CNN models, and the hand-crafted feature, such as the number or the size of nuclei, the average RGB value of nucleus, gray level co-occurrence matrix features, VGG19 features, and the number of neighbors of a nucleus~\citep{RN104}. According to the assumption that adjacent cells are more likely to interact~\citep{RN103}, the edge between the nodes can be determined via Delaunay triangulation~\citep{RN105} or the K-nearest-neighbour method~\citep{RN106}, which could evaluate whether two cells (nodes) belong to the same cluster. The cells in the same cluster are more likely to have an edge between them. Despite a good performance on clinical classification tasks, these approaches cannot work well in capturing the diagnostic and prognostic information from the surrounding micro-environments (e.g., tissues and vessels). Meanwhile, constructing cell-centered graphs highly depends on cell detection accuracy. It is notable that constructing a cell-based graph and subsequent graph computing need an excessive computational complexity. The process of utilizing GCNs in histopathological image analysis is shown in Fig.~\ref{fig4}. We outline several areas of clinical interest for GCNs in histopathology below. 

\subsubsection{Tumor segmentation}
Accurate tumor segmentation in histopathology is designed to assist pathologists for improving workflow efficiency of clinical diagnosis~\citep{RN7}. Graph-based segmentation approaches can incorporate both local and global inter-tissue-region relations to build contextualized segmentation and thus improve the overall performance. For example, SEGGINI performs semantic segmentation of images by constructing tissue-graph representation and performing weakly-supervised segmentation via node classification by using weak multiplex annotations, i.e., inexact and incomplete annotations, in prostate cancer~\citep{RN7}. In this study, they defined graph nodes by superpixels merging based on channel-wise color similarity of superpixels at higher magnification. The node attribute is determined by the spatial and morphological features of the merged node (e.g., the merged superpixel). The spatial feature is computed by normalizing superpixel centroids by the image size and the morphological feature is extracted by a pre-trained MobileNetV2~\citep{RN107}. They defined the edges by constructing a region adjacency graph (RAG)~\citep{RN108} from the spatial connectivity of superpixels. The local and global connection of tissue details creates an alternative avenue for pixel-level segmentation evaluation that draws a contrast to other conventional convolutional-based tumor segmentation approaches ~\citep{RN109, RN110, RN111}. Another study~\citep{RN112} proposed an end-to-end framework that utilizes an unsupervised pretrained CNN to extract tile features and generate dynamics superpixels for graph construction, while using GCN for predicting the final segmentation map. In this study, the dynamics superpixels can be viewed as a key bridge between CNN and the GCN model, which are generated according to the CNN feature extraction. 

\subsubsection{Cancer tissue classification}
Cancer subtype classification is crucial in clinical image analysis that can impact patient stratification, outcome assessment, and treatment development~\citep{RN6, RN113}. GCNs have been extensively studied in cancer subtype classification due to their unique ability to explore the relational features among tissue sub-regions (e.g., patches or cells). Patch-based graph construction approaches are intuitive to build a bridge between image features and graph structure. Conceptually, patches are defined as nodes and node attributes are extracted patch features, including CNN-based extracted and hand-crafted features. The edges are typically determined by the Euclidean distance of nodes. For example, the combination of ChebNet~\citep{RN14} and GraphSage~\citep{RN20} presents its usefulness for classifying lung cancer subtypes in histopathological images~\citep{RN6} via patch selection. All patches in the tissue region are grouped into multiple classes, and a portion of all clustered patches (e.g., 10\%) are randomly selected within each class. Also, a simplified graph construction process~\citep{RN6} can be useful to leverage all patch information. The global context among patches is considered while using a fully connected graph to represent the connection among nodes. Global pooling layers (e.g., global attention, max, and sum poolings) are able to generate graph representations for analyzing cancer classification. In particular, global attention pooling~\citep{RN36} provides strong interpretability to determine which nodes are relevant to the current graph-level classification tasks. In colorectal cancer histopathology, ChebNet~\citep{RN14} shows its predictive power in lymph node metastasis (LNM) prediction~\citep{RN113}. Interestingly, a combination model of a variational autoencoder and generative adversarial network (VAE-GAN)~\citep{RN114} is utilized to train as a feature extractor to decode the latent representations closer to their original data space. Further, the pixel-based graph construction could be understood as a variant of patch-based approaches. The study~\citep{RN115} developed a group quadratic graph convolutional network for breast tissue and grade classification on pixel-based graph representation. The proposed model reduces the redundant node (e.g., pixel) feature, selects superior fusion feature, and enhances the representation ability of the graph convolutional unit by the pixel-based graph analysis.

As opposed to patch-based approaches, cell-based graph construction is under a key assumption that cell-cell interactions are the most salient points of information~\citep{RN116}. A common example is to define the detected nuclei as nodes~\citep{RN104} and while the overall node attributes are aggregated by concatenating multiple types of features (see Table2). The graph edge is determined by thresholding the Euclidean distance between nodes. In addition, the cell graph convolutional network~\citep{RN103}  presents a generalized framework for grading colorectal cancer histopathological images based on the combination of GraphSage~\citep{RN20}, JK-Net~\citep{RN47}, and Diffpooling~\citep{RN38}. The edge between two nuclei is determined by a fixed distance while the maximum degree of each node is set to k corresponding to its k-nearest neighbors. Sharing a similar cell-graph construction strategy and graph component definition with~\citep{RN103}, a GIN-based~\citep{RN21} framework is designed for breast cancer subtype classification~\citep{RN117}. In addition, the clinical interpretation is provided by a cell-graph explainer that is inspired by a previous graph explainer~\citep{RN118}, a post-hoc interpretability method based on graph pruning optimization. The cell-graph explainer is able to prune the redundant graph components, such as the nodes that could not provide enough information in the decision making, and define the resulting subgraph as the explanation. Another cell graph application of cancer classification~\citep{RN119} is built on top of robust spatial filtering (RSF)~\citep{RN32}, where RSF combined with attention mechanisms to rank the graph vertices in their relative order of importance, providing visualizable results on breast cancer and prostate cancer classification. 

To leverage the advantages of patch- and cell-based graphs simultaneously, the model integration can provide additional auxiliary benefits by capturing detailed nuclei and micro-environment tissue information. A hierarchical cell-to-tissue graph neural network (HACT-Net)~\citep{RN120} is an example to consist of a low-level cell-based graph (e.g., cell-graph), a high-level patch-based graph (e.g., tissue-graph), and a hierarchical-cell-to-tissue representation for breast carcinoma subtype classification. For the cell-based graph, they defined nuclei as graph nodes that are detected by the pre-trained Hover-Net~\citep{RN50, RN121, RN122}. For the patch-based graph, they determined graph nodes and their attributes by creating non-overlapping homogeneous superpixels and their features. The edges are constructed by a region adjacency graph~\citep{RN108} using the spatial centroids of the super-pixels. Overall, such a joint analysis across histopathological scales leads to enhanced performance for cancer subtype classification.

Cancer staging and grade classification is also of clinical significance that comprises tumor tissue and nodal (e.g., tumor and lymph nodes) staging~\citep{RN116}. Patch-based graph construction strategy is commonly used in tumor staging classification in terms of graph attention~\citep{RN8}. Also, graph topological feature extraction is useful in colon cancer tumor stage prediction with well interpretation~\citep{RN116}. In particular, they utilized the Mapper~\citep{RN123} to project high-dimensional graph representation to a lower-dimensional space, summarizing higher-order architectural relationships between patch-level histological information to provide more favorable interpretations for histopathologists. Further, for liver fibrosis stage classification, the study~\citep{RN124} proposed a patch-based graph structure together with the GCN attention layer to analyze the spatial organization of the fibrosis patterns. They use the KNN algorithm to cluster the tiles to select regions of high collagen content as the centroid node for graph construction. The proposed pipeline allows for the separation of fibers in the slide into localized fibrosis patterns and the individual regions can be inspected by a pathologist~\citep{RN124}. Also, cell-based graph construction strategy is useful for cancer grade classification. For example, for prostate cancer grade classification~\citep{RN125} in tissue micro-array, GraphSage~\citep{RN20} learns the global distribution of cell nuclei, cell morphometry, and spatial features without requiring pixel-level annotation. In this study, the cell nuclei are the node of the graph while the three types of features consist of the node attribute, including the morphological feature (e.g., the area, roundness, eccentricity, convexity, orientation for each of the nucleus.), texture feature (e.g., the dissimilarity, homogeneity, energy, and ASM based on the obtained grey level co-occurrence matrix), and contrastive predictive coding features. 

\subsubsection{Survival prediction}
Survival analysis is a long-standing clinical task to determine the prognostic likelihood of patients~\citep{RN113, RN126}. Both cell- and patch-based approaches can be considered to capture survival sensitive information of patients. For instance, the graph convolutional neural network with attention learning has shown to achieve a good performance on the survival prediction in colorectal cancer~\citep{RN127}. Tumor tiles are defined as nodes and node attributes are extracted by the VGG16. Graph edges are constructed by thresholding the Euclidean distances between node attributes. After constructing the graph, they used the ChebNet~\citep{RN14} framework for survival analysis on the histopathological images. With a similar definition of graph components, another study~\citep{RN128} designed a patch-based graph construction strategy in the Euclidean space. They utilized the DeepGCN~\citep{RN85} and global attention layer to boost the survival prediction performance and provided the interpretability across five cancer types. An integrated framework~\citep{RN129} extracted morphological features from histology images using CNNs and from the constructed cell-based graph using GraphSage~\citep{RN20}, and also genomic (mutations, CNV, RNA-Seq) features using SNNs. A fusion of these deep features using the Kronecker Product is of great interest for accurate survival outcome prediction. In addition, cell-based and patch-based graphs can be further unified to allow a trade-off between efficiency and granularity~\citep{RN130}. They used GAT or prostate cancer survival prediction using WSIs. Notably, a self-supervised learning method is proposed to pretrain the model, yielding improved performance over trained-from-scratch counterparts. For cell-based graphs, they use a Mask R-CNN~\citep{RN131} for nuclei segmentation and define an eight-pixel width of the ring-like neighborhood region around each nucleus as its cytoplasm area. The nuclear morphometry features and visual texture features (intensity, gradient, and Haralick features) have made substantial contributions for both nuclear and cytoplasm region representations respectively. Despite these advances, uncertainty remains for exploring definitive roles of cell-level and patch-level characteristics with regard to overall survival likelihood of patients.

\subsubsection{Molecular biomarker prediction}
Image-based molecular biomarker prediction is promising to deepen our understanding of cancer biology across data modalities. Enormous efforts are gaining momentum to explore multiple image-to-genome associations in cancer research~\citep{RN132, RN133, RN134} . The feature-enhanced graph network (FENet)~\citep{RN12} leverages histopathological-based graph structure to predict key molecular outcomes in colon cancer. Through the spatial measurement of tumor patches, the image-to-graph transformation illustrates its unique value in predicting key genetic mutations. In particular, the use of GIN~\citep{RN21} layer and jumping knowledge structure are useful to aggregate and update the patch embedding information. Alternatively, the cell-based construction method is considerable for cancer biomarker prediction~\citep{RN135}. HoverNet~\citep{RN122} is a popular choice for nuclei segmentation to support cell graph construction. Next, the agglomerative clustering~\citep{RN136} is utilized to group spatially neighboring nuclei into clusters. These clusters can be defined as graph nodes and the node attribute is determined by the standard deviation of nuclei sizes. Meanwhile the edges are constructed by using Delauney triangulation based on the geometric coordinates of cluster centers with a maximum distance connectivity threshold. Both cell- and patch-based approaches contribute to the integration of histopathology and genome as more biological data become accessible. We recognize that graph-based models can offer an efficient means to measure the cross-modality differences, which requires careful inputs on graph construction, model layer architectures, proper design of feature extraction for achieving improved performance of molecular outcome prediction. 

Overall, Table.~\ref{tab2} summarizes the category, type of tasks, and the graph-structure construction strategies. In this chapter, we have discussed novel perspectives for computational histopathological image analysis. In particular, GCNs-based methods provide a novel perspective to consider tumor heterogeneity in histopathological image analysis. Despite multiple challenges, the evolving capacity of current graph construction strategies (edge, node, and node attributes) makes it possible to address a variety of clinical tasks using histopathological images. 

\begin{longtable}{|p{2.9cm}|p{1.2cm}|p{2cm}|p{9.5cm}|}
\caption{\label{tab2}Summary of GCNs in histopathological image analysis}\\


\hline
Method & Category & Tasks & Graph Construction\\
\hline

\endhead
\multicolumn{4}{r}{\footnotesize Continue on the next page}
\endfoot
\endlastfoot

\multirow{2}{*}{\cite{RN7}} & \multirow{2}{*}{\makecell{Patch\\based}} & \multirow{2}{*}{Segmentation} & Nodes are superpixels, and node attributes are normalizing superpixel centroids by image size and pre-trained MobileNetV2 extracted features.\\
\cline{4-4}
 & & & The graph is region adjacency graph.\\
\hline
\multirow{2}{*}{\cite{RN112}} & \multirow{2}{*}{\makecell{Patch\\based}} & \multirow{2}{*}{Segmentation} & Nodes are superpixels, and node attributes are extracted by fully-convolutional network.\\
\cline{4-4}
 & & & The edges are determined by the spatial adjacent neighbors, and edge attributes are determined by the similarity between statistical histogram features.\\
\hline
\multirow{2}{*}{\cite{RN113}} & \multirow{2}{*}{\makecell{Patch\\based}} & \multirow{2}{*}{\makecell{LNM\\ Prediction}} & Nodes are image patches, and node attributes are extracted features and closer the feature space to the original one by VAE-GAN.
\\
\cline{4-4}
 & & & The edges are determined by calculating the Euclidean distance of node attributes\\
\hline
\multirow{2}{*}{\cite{RN6}} & \multirow{2}{*}{\makecell{Patch\\based}} & \multirow{2}{*}{\makecell{Cancer\\ Type/subtype \\classification}} & Nodes are image patches, and node attributes are extracted by DenseNet.
\\
\cline{4-4}
 & & & The graph is fully connected graph.\newline\\
\hline

\multirow{2}{*}{\cite{RN116}} & \multirow{2}{*}{\makecell{Patch\\based}} & \multirow{2}{*}{\makecell{Tumor Stage\\Prediction}} & Nodes are image patches, and node attributes are extracted features of images patch.
\\
\cline{4-4}
 & & & The edges are determined by the spatial relationship between nodes.\\
\hline
\multirow{2}{*}{\cite{RN8}} & \multirow{2}{*}{\makecell{Patch\\based}} & \multirow{2}{*}{\makecell{TNM Stage\\Prediction}} & Nodes are image patches, and node attributes are texture feature extraction.
\\
\cline{4-4}
 & & & The edges are determined by the connectivity between nodes and their KNN neighbors in a fixed threshold.\\
\hline
\multirow{2}{*}{\cite{RN115}} & \multirow{2}{*}{\makecell{Pixel\\based}} & \multirow{2}{*}{\makecell{Cancer tissue \\and grade \\classification}} & Nodes are image pixels, and node attributes are extracted by ResNet18.
\\
\cline{4-4}
 & & & The edges are determined by the connectivity between nodes and their KNN neighbors.\\
\hline
\multirow{2}{*}{\cite{RN124}} & \multirow{2}{*}{\makecell{Patch\\based}} & \multirow{2}{*}{\makecell{Cancer tissue \\and grade \\classification}} & Nodes are image patches, and node attributes are extracted by ResNet18.
\\
\cline{4-4}
 & & & The edges are determined by the connection between patch and the centroid node of each dense collagen region. The edge attribute is encoded by the Euclidean distance from each tile to its corresponding centroid.\\
\hline

\multirow{2}{*}{\cite{RN125}} & \multirow{2}{*}{\makecell{Cell\\based}} & \multirow{2}{*}{\makecell{TMA\\Grade \\classification}} & Nodes are nuclei, and node attributes are morphological, texture, and contrastive predictive coding features.
\\
\cline{4-4}
 & & & The edge is determined by nodes and their KNN neighbors.\\
\hline
\newpage
\multirow{2}{*}{\cite{RN104}} & \multirow{2}{*}{\makecell{Cell\\based}} & \multirow{2}{*}{\makecell{Cancer \\type/subtype\\classification}} & Nodes are nuclei, and node attributes are concatenated by multiple types of features, including average RGB value, gray level co-occurrence matrix features, VGG19 features, and the number of neighbors of a nucleus. 
\\
\cline{4-4}
 & & & The edge connections are determined by thresholding the Euclidean distance between nuclei.\\
\hline
\multirow{2}{*}{\cite{RN103}} & \multirow{2}{*}{\makecell{Cell\\based}} & \multirow{2}{*}{\makecell{Cancer \\type/subtype\\classification}} & Nodes are nuclei, and node attributes are 16 hand-craft features and 17 nuclear descriptors.
\\
\cline{4-4}
 & & & The edges are determined by the connectivity between nodes and their KNN neighbors.\\
\hline

\multirow{2}{*}{\cite{RN117}} & \multirow{2}{*}{\makecell{Cell\\based}} & \multirow{2}{*}{\makecell{Cancer \\type/subtype\\classification}} & Nodes are nuclei, and node attributes are 16 hand-crafted features.
\\
\cline{4-4}
 & & & The edges are determined by thresholding the kNN graph by removing edges longer than a specified distance.\\
\hline

\multirow{2}{*}{\cite{RN119}} & \multirow{2}{*}{\makecell{Cell\\based}} & \multirow{2}{*}{\makecell{Cancer \\type/subtype\\classification}} & Nodes are nuclei, and node attributes are concatenated edge and vertex features of nodes.
\\
 & & & The edges are determined by the Euclidean distance between nuclei\\
\hline

\multirow{4}{*}{\cite{RN120}} & \multirow{2}{*}{\makecell{Patch\\based}} & \multirow{4}{*}{\makecell{Cancer \\type/subtype\\classification}} & Nodes are superpixel, and node attributes are features of superpixels.
\\
\cline{4-4}
& & & The graph is region adjacency graph.
\\
\cline{2-2}
\cline{4-4}
& \multirow{2}{*}{\makecell{Cell\\based}}& & Nodes are nuclei, and node attributes are hand-craft features.\\
\cline{4-4}
 & & & The edges are determined by the connectivity between nodes and their KNN neighbors.\\
\hline
\multirow{2}{*}{\cite{RN127}} & \multirow{2}{*}{\makecell{Patch\\based}} & \multirow{2}{*}{\makecell{Survival\\prediction}} & Nodes are patches, and node attributes are extracted by VGG16.
\\
\cline{4-4}
 & & & The edges are determined by the Euclidean distances between node attributes.\\
\hline
\multirow{2}{*}{\cite{RN128}} & \multirow{2}{*}{\makecell{Patch\\based}} & \multirow{2}{*}{\makecell{Survival\\prediction}} & Nodes are patches, and node attributes are extracted by ResNet50.
\\
\cline{4-4}
 & & & The edges are determined by the Euclidean distances between node coordinates.\\
\hline
\multirow{2}{*}{\cite{RN129}} & \multirow{2}{*}{\makecell{Cell\\based}} & \multirow{2}{*}{\makecell{Survival\\prediction}} & Nodes are patches, and node attributes are hand-craft and contrastive predictive coding features.
\\
\cline{4-4}
 & & & The edges are determined by the connectivity between nodes and their KNN neighbors.\\
\hline
\multirow{4}{*}{\cite{RN130}} & \multirow{2}{*}{\makecell{Patch\\based}} & \multirow{4}{*}{\makecell{Cancer \\type/subtype\\classification}} & Nodes are patches, and node attributes are image features and cell-based graph representation.
\\
\cline{4-4}
& & & The edges are determined by the connectivity between nodes and their KNN neighbors.
\\
\cline{2-2}
\cline{4-4}
& \multirow{2}{*}{\makecell{Cell\\based}}& & Nodes are nuclei, and node attributes are nuclear morphometry features and imaging features (including intensity, gradient and Haralick features.\\
\cline{4-4}
 & & & The edges are determined by the connectivity between nodes and their KNN neighbors.\\
\hline
\multirow{2}{*}{\cite{RN12}} & \multirow{2}{*}{\makecell{Patch\\based}} & \multirow{2}{*}{\makecell{Biomarker\\Prediction}} & Nodes are patches, and node attributes are extracted by ResNet18.
\\
\cline{4-4}
 & & & The edges are determined by thresholding the Euclidean distance between node coordinates.\\
 \hline
 \multirow{2}{*}{\cite{RN135}} & \multirow{2}{*}{\makecell{Cell\\based}} & \multirow{2}{*}{\makecell{Biomarker\\Prediction}} & Nodes are the geometric center of nuclei cluster, and node attributes are determined by the count of the six nuclei types and the standard deviation of nuclear sizes.
\\
\cline{4-4}
 & & & The edges are determined by the Delauney triangulation between cluster center with a maximum distance threshold.\\
 \hline
\end{longtable}

\section{Other image-based applications}
GCNs have demonstrated their analytical ability in alternative medical image disciplines to facilitate structural analysis of disease diagnosis (e.g., eye disease and skin lesion), surgery scene understanding, and Bone Age Assessment. For instance, GCNs have been studied in dermatology and eye-related diseases, involving retinal, fundus, and fluorescein angiography (FA) images~\citep{RN126,RN137, RN138, RN139}. Similar to radiological and histopathological images, patch-based graph construction strategies are widely used in the above image domains. GCNs have shown to be valuable to learn the vessel shape structures and local appearance for vessel segmentation in retinal images~\citep{RN137}. Also, GCNs were applied to the artery and vein classification by using both fundus images and corresponding fluorescein angiography (FA) images~\citep{RN138}. With a designed graph U-Nets architecture~\citep{RN40}, the high-level connectivity of vascular structures can be learned from node clustering in the node pooling layers. In addition, a study~\citep{RN140} proposed a framework that combines the CNN and ResGCN~\citep{RN85} model to enhance the segmentation performance of fetal head on ultrasound images. Furthermore, GCNs show their power in differential diagnosis of skin conditions using clinical images~\citep{RN139}. This problem is formulated as a multi-label classification task over 80 conditions when only incomplete image labels are available. The label incompleteness is addressed by combining a classification network with a graph convolutional network that characterizes label co-occurrence~\citep{RN139}. Each clinical image is defined as a graph node and the connectivity between two nodes is determined by domain knowledge of skin condition by board-certified dermatologists. It is noteworthy that edge connection is made by inputs from human experts that two dermatologists provide overlapped differential diagnoses groups and connect an edge when two labels appear in at least one differential group by both dermatologists. In addition, a cell-based graph analysis~\citep{RN126} combines multiple types of GCNs with graph poolings, including GIN~\citep{RN20}, GraphSage~\citep{RN21}, and GCN~\citep{RN16} for survival prediction of gastric cancer using immunohistochemistry (mIHC) images. The graph nodes are determined by six antibodies of PanCK, CD8, CD68, CD163, Foxp3, and PD-L1, which were used as annotation indicators for six different types of cells. The node attributes are determined by cell locations, types, and morphological features. The edges are constructed by the maximum effective distance between immune and tumor cells, which is equivalent to 40 pixels in the magnification of this study. Furthermore, a surgery scene graph analysis~\citep{RN141} utilized GraphSage to predict surgical interactions between instruments and surgical regions of interest. In addition, GCNs provide the potential for automatic bone age assessment and ROI score prediction on hand radiograph~\citep{RN142}. This study proposed an anatomy-based group convolution block to predict the ROI scores by processing the local features of ROIs. Also, they presented a dual graph-based attention module to compute the patient-specific attention and context attention for ROI score prediction. 

\begin{longtable}{|p{2.7cm}|p{2.2cm}|p{2cm}|p{9cm}|}
\caption{\label{tab3}Summary of GCNs in other image analysis}\\


\hline
Method & Image Type & Tasks & Graph Construction\\
\hline

\endhead
\multicolumn{4}{r}{\footnotesize Continue on the next page}
\endfoot
\endlastfoot

\multirow{2}{*}{\cite{RN126}} & \multirow{2}{*}{\makecell{Immuno-\\histochemistry\\(mIHC) images}} & \multirow{2}{*}{\makecell{Survival\\prediction}} & Nodes are cell, and node attributes are determined by the cell locations, optical features of stained cells, and morphology features.\\
\cline{4-4}
 & & & The edges are determined by the Euclidean distances between nodes, and edge attributes is calculated by the follow equation:
 \begin{equation}
     \frac{40}{disttance between cells}
 \end{equation}
It will be set to 0 while no interaction between nodes.
 \\
\hline
\multirow{2}{*}{\cite{RN137}} & \multirow{2}{*}{Retinal image} & \multirow{2}{*}{\makecell{Vessel\\segmentation}} & Nodes are pixels with maximum vessel probability, and node attributes are extracted by CNN.\\
\cline{4-4}
 & & & The edges are determined by thresholding the Geodesic distance between nodes.
 \\
\hline
\newpage
\multirow{2}{*}{\cite{RN138}} & \multirow{2}{*}{\makecell{Fundus images\\and \\corresponding\\fluorescein \\angiography \\images}} & \multirow{2}{*}{\makecell{The artery \\ and vein \\classification}} & Nodes are vessel pixels with in $N\times N$ local patches, and node attributes are extracted by graph U-nets.
\newline
\\
\cline{4-4}
 & & & The edges are constructed with existing vessel pixels within an $N\times N$ local patch.
 \newline
 \\
\hline
\multirow{2}{*}{\cite{RN140}} & \multirow{2}{*}{\makecell{Fundus images \\ ultrasound \\images}} & \multirow{2}{*}{\makecell{Optic disc \\and cup \\segmentation\\ and fetal head \\segmentation}} & Nodes are the object boundaries that are divided into N vertices with the same interval, and node attributes are extracted by CNN.\\
\cline{4-4}
 & & & The edges are determined by every two consecutive vertices on the boundary and the center vertex are connected to form a triangle. 
\\
\hline
\multirow{2}{*}{\cite{RN139}} & \multirow{2}{*}{\makecell{Skin clinical\\images}} & \multirow{2}{*}{\makecell{Skin condition\\classification}} & Nodes are images, and node attributes are extracted by CNN.\\
\cline{4-4}
 & & & The edges are connected when two labels appear in at least one differential group by both dermatologists, and edge attributes is calculated by the follow equation:
 \begin{equation}
     \frac{C(i,j)}{C(i) + C(j)}
 \end{equation}
$C(i,j)$ is the number of images have two label at same time. $C(i)$ or ${C(j)}$ is the number of images in class $i$ or $j$.
\\
\hline

\multirow{2}{*}{\cite{RN141}} & \multirow{2}{*}{\makecell{Surgery scene\\images}} & \multirow{2}{*}{\makecell{Surgery scene\\understanding}} & Nodes are surgical tools and defective tissues , and node attributes are extracted by CNN with label smoothing.\\
\cline{4-4}
 & & & The edges are determined by the interactions between nodes.
 \\
\hline
\multirow{2}{*}{\cite{RN142}} & \multirow{2}{*}{\makecell{Hand radiograph}} & \multirow{2}{*}{\makecell{Bone age \\assessment}} & Nodes are ROIs , and node attributes are extracted by CNN.\\
\cline{4-4}
 & & & The edges are determined by the natural connections among ROIs and the full connections among the ROIs in the same anatomy group.
 \\
\hline

\end{longtable}

\section{Discussion and future direction}
The rapid growth of GCNs~\citep{RN12, RN113, RN127} and their extensions have been increasingly utilized for processing, integrating, and analyzing multi-modality medical imaging and other types of biological data. We here discuss several future research directions and common challenges to advance the research in medical image analysis and related research fields. We particularly outline key aspects of importance, including GCN model interpretation, the value of pre-training model, evaluation pipeline, large-scale benchmark, and emerging technical insights.

\subsection{Interpretability}
The interpretation of GCNs is of heightened interest to make the outcome understandable, ensure model validity, and enhance clinical relevance. In our focus, a well-designed interpretation framework of GCNs is expected to provide the explanation and visualization for both image-wise and graph components understanding. Such an interpretative ability can be highly attractive to clinicians in the process of diagnosing regions of interest in histopathology, enabling an understanding of spatial and regional interactions from graph structures~\citep{RN12}. As demonstrated, three metrics are useful to design and understand the interpretation capability of GCNs~\citep{RN143}: (1) Fidelity refers to the importance of classification as measured by the impact of node attributes, (2) Contrastively points to the significance with respect to different classes, and (3) Sparsity reflects the sparseness level on a graph. These metrics can help generate and measure the valuable heat maps of graph nodes given their attributes. Representative approaches include gradient class activation mapping (Grad-CAM), contrastive excitation backpropagation (c-EB), and contrastive gradient (CG)~\citep{RN144}. Further, we recognize that emerging studies have explored the specified interpretation strategy for GCNs~\citep{RN117, RN118}. For instance, an ROI-selection pooling layer (R-pool)~\citep{RN143} highlights the node importance for predicting neurological disorders by removing noisy nodes to realize a dimension reduction of the entire graph. Rather than node-level feature interpretation, additional efforts will be greatly needed on interpreting the relational information in graphs. GnnExplainer~\citep{RN118} is an example to leverage the recursive neighborhood-aggregation scheme to identify graph pathways as well as node feature information passing along the edges. The design of GnnExplainer is appealing to visualize the detailed cell-graph structure and provide class-specific interpretation for breast cancer~\citep{RN117}. As a result, we strongly emphasize that the interpretation process considers an in-depth joint understanding of the clinical task, graph model architecture, and model performance. 

\subsection{Model pretraining}
Pretraining GCNs aims to train a model on the tasks with a sufficient amount of data and labels and finetune the model into downstream tasks. Pre-trained GCNs can serve as a foundation model to improve the generalization power when the size of the training set is often limited in medical imaging~\citep{RN145}. The pretraining workflow of GCNs typically includes the model training rules, hyper-parameter settings, and constructed-graph augmentation strategies. A key pretraining scheme for a graph-level task is to reconstruct the vertex adjacency information (e.g. GraphSAGE) without hurting intrinsic structural information~\citep{RN146}. We offer several compelling directions of pretraining strategy to improve GCNs model robustness and their utility in different tasks. First, the graph-wise augmentation strategies have a large room to facilitate the pretraining of graphs. For instance, the out-of-distribution samples can be analyzed via node-level and graph-level augmentations~\citep{RN145, RN146}. Second, exploring label-efficient models (e.g., unsupervised or self-supervised learning) in conjunction with pretraining strategies~\citep{RN146} could greatly alleviate the labeling shortage. Notable studies~\citep{RN145, RN146, RN147} have achieved good performance in downstream tasks while leveraging the graph-based pretraining strategies. Considering the above directions, a self-supervised learning framework for GCNs pretraining~\citep{RN146} demonstrates that graph-wise augmentation strategies are useful to address the graph data heterogeneity. The pretraining is performed through maximizing the agreement between two augmented views of the same graph via performing node dropping, edge perturbation, attribute masking, and subgraph selection. Notably, only a small partition of graph components will be changed, meanwhile the semantic meaning of the graph has been preserved. Such a strategy brings graph data diversity that is greatly needed for building robust pre-trained GCN models. Taken together, the research on pretraining GCNs and their practical impact is only to start and will continue to make progress on downstream image-related clinical tasks.

\subsection{Evaluation of graph construction strategies }
The evaluation of graph construction in medical imaging is vital because the associated graph construction could significantly affect the model performance and the interpretation of outcomes. The general graph construction methods used ROIs (e.g., image patches or brain neurons) as graph nodes and the node attributes are obtained by standard feature extractors (e.g., ResNet18). In addition, edges represent the connections between nodes which could be determined by the Euclidean distance between node features, or the connections between ROIs which are determined by patch coordinates or the actual neural fiber connections. Currently, graph construction strategies are applied in different tasks and a generalized graph construction evaluation strategy is not explicitly developed yet. It is even more difficult to determine which kind of graph construction is generalizable for task-specific medical image analysis because of various datasets and graph construction metrics. Also, developing a generalized graph construction evaluation strategy is necessary for GCNs to better process medical image data across multiple modalities because the model performance is highly related to the quality of constructed graph-structured data. The benchmarking framework~\citep{RN158, RN149} has rigorously evaluated the performance of graph neural networks on medium-scale datasets and demonstrates its usefulness for analyzing message-passing capability in GCNs. Also, a comparison strategy among multiple GCNs~\citep{RN133} can address the issues of reproducibility and replicability. Following the graph evaluation~\citep{RN132, RN133}, we need to define statistical distinctions to ensure the performance of GCNs. For example, it is helpful for model training and human understanding if the graph structure and feature distribution differences between positive and negative patient samples are significantly different.

\subsection{Real-world large-scale graph benchmark}
Despite the remarkable effort on standardization of medical imaging cohorts, the high-quality, large-scale graph-defined benchmark has not been readily available for AI model evaluation, especially in medical image analysis. Open Graph Benchmark (OGB) exemplifies the initiative that contains a diverse set of real-world benchmark datasets (e.g., protein, drug, and molecular elements) to facilitate scalable and reproducible graph machine learning research~\citep{RN150}. The number of graphs and nodes in each graph are both massive in OGB. Even small-scale OGB graphs can have more than 100 thousand nodes or more than 1 million edges. This comprehensive dataset in various domains can be viewed as a baseline to support the GCNs’ development and comparison. Related works have been explored on graphs including a chemistry dataset with 2 million graphs and a biology dataset with 395K graphs~\citep{RN145}. As seen in OGB development, there are challenges to collecting and processing suitable medical image datasets and constructing meaningful graphs following the image-to-graph transformation. First, we need to collect a large number of medical images across multiple centers to ensure data diversity. It is also essential to provide detailed annotation information for collected datasets on the image-level region of interest. Second, graph-wise statistics is important to allow measurement of graph-level dynamics. Notable graph metrics~\citep{RN151}, such as the average node degree, clustering coefficient, closeness centrality, and  betweenness centrality, can be used to assess graph characteristics and help determine unique graph structures. For instance, the average node degree calculates the average degree of the neighborhood of each node to delineate the connectivity between nodes to their neighbors. The clustering coefficient measures how many nodes in the graph tend to cluster together. Closeness centrality highlights nodes that can easily access other nodes. Third, we must carefully design image-graph components, such as the definition of graph nodes in different types of graphs that are vital to downstream clinical tasks. Finally, the real impact of pre-trained foundation models on large-scale graph-wise datasets still needs to be explored. While using pretraining GCNs to improve data-efficiency issues in medical image analysis, the models can be well-trained on the large-scale graph-wise dataset and adapt into specific tasks, even with a limited size of downstream data.

\subsection{Technological advancements}
The rapid development of deep learning is bringing novel perspectives to address the challenges of graph-based image analysis. The transformer architecture~\citep{RN23}, emphasizing the use of a self-attention mechanism to explore long-range sequential knowledge, emerges to improve the model performance in a variety of natural language processing (NLP)~\citep{RN152} and computer vision tasks~\citep{RN153, RN154}. A graph-wise transformer can be effectively considered to capture both local and global contexts, thus holding the promise to overcome the limitation of spatial-temporal graph convolutions. For example, graph convolutional skeleton transformers integrate both dynamical attention and global context, as well as local topology structure in GCNs~\citep{RN155} while the spatial transformer attention module discovers the global correlations between the bone-connected and the approximated connected joints of graph topology. In medical image analysis, the combination of GCNs and transformer models can be favored to process 3D MRI sequences to boost the model prediction performance, where GCNs explore the topological features while Transformers could model the temporal relationship among MRI sequences. In the meantime, self-supervised learning strategy is emerging in graph-driven analysis with limited availability of imaging data. Notably, self-supervised learning (SSL) provides a means to pretrain a model with unlabeled data, followed by fine-tuning the model for a downstream task with limited annotations~\citep{RN156}. Contrastive learning (CL), as a particular variant of SSL, introduces a contrastive loss to enforce representations to be close for similar pairs and far for dissimilar pairs~\citep{RN156, RN157}. Another technique to address the limitation of data labeling is the advent of self-training learning to generate pseudo-label for model retraining and optimization~\citep{RN158}. A self-training method for MRI segmentation has shown the potential solution for cross-scanner and cross-center data analytical tasks~\citep{RN158}. Also, the teacher-student framework is another type of self-training, which trains a good teacher model with labeled data to annotate the unlabeled data, and finally, the labeled data and data with pseudo-labels can jointly train a student model~\citep{RN131}. Overall, both self-supervised learning and self-training strategies can be utilized in GCNs model training to potentially improve the model performance and overcome the annotation and data scale challenges.

\section{Conclusion}
We have witnessed a growing trend of graph convolutional networks applied to medical image analysis over the past few years. The convergence of GCNs, medical imaging data, and other clinical data, brings advances into outcome interpretation, disease understanding, and novel insights into data-driven model assessment. These breakthroughs, together with data fusion ability, local and global feature inference, and model training efficiency, lead to a wide range of downstream applications across clinical imaging fields. Nevertheless, the development of benchmark graph-based medical datasets is yet to be established. Consistency and validity of graph construction strategy in medical imaging are greatly needed in future research. Recent technological advances can be considered to enhance and optimize GCNs in addressing challenging problems. We hope that the gleaned insights of this review will serve as a guideline for researchers on graph-driven deep learning across medical imaging disciplines and will inspire continued efforts on data-driven biomedical research and healthcare applications.


\bibliographystyle{model2-names.bst}\biboptions{authoryear}
\bibliography{refs}



\end{document}